%% file: Main.tex
\documentclass[journal]{IEEEtran}

\usepackage[utf8]{inputenc}
\usepackage[T1]{fontenc}

\usepackage[pass]{geometry}

\usepackage{amsfonts}
\usepackage{amsmath}
\usepackage{amssymb}
\usepackage{graphicx}
\graphicspath{{./}{../../docs/figures/}}
\usepackage{cite}
\usepackage{array}
\usepackage{booktabs}
\usepackage{balance}
\usepackage[final,expansion=false]{microtype}
\usepackage{titlesec}
\usepackage{tabularx}
\usepackage[table,xcdraw]{xcolor}
\usepackage{multirow}
\usepackage[normalem]{ulem}
\useunder{\uline}{\ul}{}

\usepackage{pifont}

\usepackage{scalerel}
\usepackage{tikz}

\usepackage[acronym,shortcuts]{glossaries-extra}
\glssetcategoryattribute{acronym}{nohyper}{true}
\setabbreviationstyle[acronym]{long-short}

\newacronym{cps}{CPS}{Cyber-Physical System}
\newacronym{fdia}{FDIA}{False Data Injection Attack}
\newacronym{ied}{IED}{Intelligent Electronic Device}
\newacronym{iiot}{IIoT}{Industrial IoT}
\newacronym{iot}{IoT}{Internet of Things}
\newacronym{nist}{NIST}{National Institute of Standards and Technology}
\newacronym{pa}{PA}{Policy Administrator}
\newacronym{pe}{PE}{Policy Engine}
\newacronym{pdp}{PDP}{Policy Decision Point}
\newacronym{pep}{PEP}{Policy Enforcement Point}
\newacronym{rtu}{RTU}{Remote Terminal Unit}
\newacronym{sa-zt}{SA-ZT}{Safety-Aware Zero Trust}
\newacronym{se}{SE}{Safety Engine}
\newacronym{tb}{TB}{Telemetry Broker}
\newacronym{wls}{WLS}{Weighted Least-Squares}
\newacronym{zt}{ZT}{Zero Trust}
\newacronym{zta}{ZTA}{Zero Trust Architecture}

\usepackage{tcolorbox}

\usepackage[hidelinks]{hyperref}

\title{Safety-Aware Zero Trust Enforcement for \\IoT and Cyber-Physical Systems}
\author{
    \IEEEauthorblockN{
    Alessandro Lotto
    Alessandro Brighente,
    and Mauro Conti
    }\\
    \IEEEauthorblockA{
    University of Padua / Department of Mathematics\\
    Padua, IT\\
    \{alessandro.lotto\}@phd.unipd.it\\
    \{alessandro.brighente, mauro.conti\}@unipd.it
    }
}

\begin{document}

\maketitle

\begin{abstract}
Zero Trust (ZT) replaces the implicit trust assumptions of perimeter-based security models with explicit, continuous, and context-aware authorization.
This shift is especially relevant to IoT and cyber-physical systems, which comprise heterogeneous, long-lived, and remotely connected components.
However, their coupling to physical processes makes ZT adoption particularly challenging: restricting a suspicious component may reduce cyber exposure while also removing telemetry or control capabilities needed for operation.
Existing work primarily models the physical harm caused by attacks, while giving less attention to the consequences introduced by enforcement itself.

We introduce Safety-Aware Zero Trust (SA-ZT), a framework that treats restriction-induced physical consequences as policy inputs.
To ground its principles, we map the NIST ZT tenets to nine IoT/CPS convergence strains, distinguish IoT-amplified challenges from those arising specifically from cyber-physical coupling, and derive the operational requirements that SA-ZT should satisfy.
SA-ZT extends NIST ZT Architecture with a Safety Engine and a Telemetry Broker.
The Safety Engine selects among admissible responses by jointly considering residual cyber risk and restriction-induced consequences, while the Telemetry Broker mediates raw telemetry visibility and estimator influence.
Together with command-side enforcement, these entities separate raw visibility, automated influence, and state-changing authority, allowing observations to remain available for monitoring while their influence on automated control is constrained.
We illustrate how SA-ZT makes the relationships among cyber containment, telemetry visibility and influence, restriction-induced physical consequences, and authorization timing explicit, in an IEEE 30-bus case study under false-data-injection attack.
This demonstrates an implementable and inspectable representation of cyber-physical enforcement trade-offs.\\

\end{abstract}

\begin{IEEEkeywords}
Zero Trust, Cyber-Physical Systems, IoT, Physical Safety, Safety-Aware Enforcement.
\end{IEEEkeywords}

\input{Sections/1-Introduction}
\input{Sections/2-Background}

\input{Sections/3-RelatedWork}
\input{Sections/4-ZT_strains}
\input{Sections/5-SafetyAware_ZT}
\input{Sections/6-IEEE30-Model}
\input{Sections/7-Evaluation}
\input{Sections/8-Discussion}
\input{Sections/Conclusion}

\bibliographystyle{IEEEtran}
\bibliography{Bibliography}

\appendices
\input{Sections/Appendix_revised}

\end{document}

%% file: Sections/1-Introduction.tex
\section{Introduction} \label{sec:introduction}
Traditional perimeter-based security models infer trust from location within a defined network boundary.
The growth of \ac{iot}, cloud computing, remote access, and distributed infrastructures has made that boundary increasingly difficult to define and defend~\cite{NIST800207,CISA2023ZTMM,Theory_ZT,Survey_ZT,SYED2022}.
The \ac{zt} security model emerged in response, requiring explicit authorization based on the requesting subject, the protected resource, and contextual evidence available at decision time~\cite{KINDERVAG2010,NIST800207,ComprehensiveReview_ZT,Survey_ZT,Theory_ZT}.
Extending this model to the \ac{iot}, \ac{iiot}, and \acp{cps} is especially important because these environments contain heterogeneous, long-lived, resource-constrained, and potentially exposed devices whose compromise may disrupt critical operations or cause physical harm~\cite{Humayed,Giraldo,Ding,Roman2013,Sicari2015,Mosenia2017,Neshenko2019,HaddadPajouh}.
This extension also changes the consequences of authorization.
Network-level \acp{pep}, protecting heterogeneous or legacy resources, often support only coarse actions, such as allowing, blocking, segmenting, or terminating communication~\cite{FEDERICI2023,CP-ZTA}.
In enterprise IT, denial is generally treated as a conservative response to insufficient trust, even when it temporarily impairs availability~\cite{SaltzerSchroeder1975,NIST800207}.
In a \ac{cps}, however, communication availability is often operationally critical, and blocking a suspicious connection may impair observation or control of the physical process~\cite{Humayed,Giraldo,Ding}.
Restricting it can therefore reduce an attacker's opportunities while also removing sensing or actuation capabilities needed for legitimate operation.
Residual cyber exposure and the consequences introduced by restriction must consequently be represented as distinct quantities~\cite{ABUR2004}.

Existing \ac{zt} research establishes principles for continuous authorization, trust evaluation, micro-segmentation, and dynamic policy enforcement~\cite{ComprehensiveReview_ZT,Survey_ZT,Theory_ZT,Critical_ZT,SystematicReview_ZT}.
Convergence studies adapt these principles to constrained endpoints, heterogeneous protocols, and limited administrative control, while recent proposals incorporate the estimated physical impact of compromise into trust or policy evaluation~\cite{Verify_and_trust,Converging_ZT,ZT_ContextIoT,CP-ZTA}.
Nevertheless, enforcement remains predominantly connection- or flow-oriented, and physical information is used mainly to determine whether access should be restricted.
Existing approaches provide less guidance on which capabilities should be withdrawn, which information should remain available, or whether the defensive restriction introduces a separate physical consequence. Addressing this gap requires both a systematic account of how IoT/CPS properties strain ZT assumptions and enforcement semantics that distinguish observation from automated influence and state-changing authority.

In this paper, we investigate \ac{zt}-\ac{cps} convergence explicitly considering the physical consequences in the decision process.
Our investigation is guided by three research questions.
\begin{itemize}
    \item \textbf{RQ1}: How do \ac{iot} and \ac{cps} characteristics strain the NIST \ac{zta} tenets, and which resulting strains amplify established \ac{zt} challenges rather than arise specifically from cyber-physical coupling?

    \item \textbf{RQ2}: What operational requirements are needed to address these strains while preserving the implementability and operational validity of \ac{zt} controls?
    
    \item \textbf{RQ3}: How can \ac{zt}--\ac{cps} convergence constrain a suspicious device's ability to influence the physical process without unnecessarily suppressing telemetry required for operational awareness?
\end{itemize}

To answer RQ1, we map the \ac{nist} \ac{zta} tenets to nine convergence strains and distinguish \ac{iot}-amplified challenges from those arising specifically from cyber-physical coupling.
To answer RQ2, we derive the corresponding operational requirements for deployable controls, timely and evidence-valid decisions, and physical-consequence-aware enforcement.
These analyses show that authorization must determine not only whether a component should be restricted, but which capabilities may remain available under the current operating conditions.
To answer RQ3, we introduce \ac{sa-zt}, which, to the best of our knowledge, is the first \ac{zt} framework to treat restriction-induced physical consequences as explicit policy inputs.
\ac{sa-zt} extends \ac{zta} with a \emph{\ac{se}} and a \emph{\ac{tb}}.
The \ac{se} ranks graduated responses that satisfy declared admissibility conditions by keeping residual cyber risk, operational degradation, restriction-induced physical consequences, and reconfiguration costs distinct.
The \ac{tb} mediates raw telemetry visibility and estimator influence, while command-side enforcement constrains state-changing authority.
This capability separation permits selected telemetry to remain available for monitoring even when its automated influence or the associated command authority is restricted.
We provide an illustrative instantiation on the IEEE 30-bus benchmark~\cite{MATPOWER} under \ac{fdia} scenarios, in which manipulated measurements can bias state estimation and automated control~\cite{LIU2011}. The implementation maps the framework to causal telemetry and command decisions, declared validity checks, bounded response selection, fallback behavior, and explicit authorization timing.
The evaluation exposes containment-information trade-offs, confirms that restricting command authority can itself produce modeled physical consequences, and shows that dynamic response information can affect a subset of decisions while activation timing influences recovery behavior.
These findings demonstrate that cyber-physical enforcement trade-offs can be represented, implemented, and inspected.
Nevertheless, they do not establish universal safety or consistent control-performance improvement.

The remainder of the paper is organized as follows.
Sec.~\ref{sec:background} introduces the \ac{nist} \ac{zta} tenets and reference model, and Sec.~\ref{sec:related-work} reviews related work.
Sec.~\ref{sec:zt-iot-strains} analyzes the nine convergence strains and derives the operational requirements, addressing RQ1 and RQ2.
Sec.~\ref{sec:general-safety-engine} presents the general \ac{sa-zt} framework, addressing RQ3, while Sec.~\ref{sec:ieee30-safety-engine-instantiation} instantiates it on the IEEE 30-bus system.
Sec.~\ref{sec:ieee30-experimental-evaluation} presents the experimental evaluation.
Sec.~\ref{sec:discussion} discusses the results, answers the research questions, and examines the implications and limitations.
Finally, Sec.~\ref{sec:conclusion} concludes the paper.

%% file: Sections/2-Background.tex
\section{Background: Zero Trust Architecture}   \label{sec:background}
\ac{zt} is a resource-centric security model in which authorization is explicit, dynamic, and granular.
Trust is not inferred from network location or asset ownership, nor carried forward automatically from a previous authentication or authorization decision.
Instead, each request is evaluated before access is granted~\cite{KINDERVAG2010,NIST800207}.
A \ac{zta} implements this security model.
\ac{nist} expresses it through seven foundational tenets, summarized in Table~\ref{tab:NIST-Tenets}~\cite{NIST800207}.
Collectively, these tenets shift security from protecting a fixed network boundary to continuously controlling interactions between subjects and individual resources.

The \ac{nist} reference model centers on three core logical components.
The \ac{pe} grants, denies, or revokes access by evaluating enterprise policy against contextual information.
This includes identity, device posture, asset state, behavioral signals, and threat intelligence.
The \ac{pa} executes this decision by configuring the relevant \acp{pep}.
The \ac{pep}, positioned on the data path, enables, monitors, and terminates connections between subjects and protected resources.
Together, the \ac{pe} and \ac{pa} form the \ac{pdp}, while the \ac{pep} mediates the resulting subject-resource communication~\cite{NIST800207}.

\input{Tables/Tenets}



%% file: Tables/Tenets.tex
\begin{table}[!h]
\centering
\caption{NIST Zero-Trust Architecture tenets~\cite{NIST800207}}
\label{tab:NIST-Tenets}
\renewcommand{\arraystretch}{1.25}
\resizebox{\columnwidth}{!}{%
\begin{tabular}{c|l}
\rowcolor[HTML]{C0C0C0} 
\begin{tabular}[c]{@{}c@{}}\textbf{ NIST }\\\textbf{Tenet}\end{tabular} & \multicolumn{1}{c}{\cellcolor[HTML]{C0C0C0}\textbf{Zero Trust Principle}}                                                                           \\ \hline
\textit{T1}         & All data sources and computing services are treated as resources                                                                            \\
\rowcolor[HTML]{EFEFEF} 
\textit{T2}         & All communication must be secured regardless of network location                                                                            \\
\textit{T3}         & \begin{tabular}[c]{@{}l@{}}Access to each resource is granted on a per-session\\ and least-privilege basis\end{tabular}                     \\
\rowcolor[HTML]{EFEFEF} 
\textit{T4} &
  \begin{tabular}[c]{@{}l@{}}Dynamic policies account for the observable state of the\\ subject identity, application or service, and requesting asset,\\ together with relevant behavioral and environmental attributes\end{tabular} \\
\textit{T5}         & \begin{tabular}[c]{@{}l@{}}Continuous monitoring of the integrity and security posture\\ of all owned and associated assets\end{tabular}    \\
\rowcolor[HTML]{EFEFEF} 
\textit{T6}         & \begin{tabular}[c]{@{}l@{}}Authentication and authorization before access while\\ continually reassessing ongoing interactions\end{tabular} \\
\textit{T7}         & \begin{tabular}[c]{@{}l@{}}Collect information about assets, network infrastructure,\\ communications, and access requests\end{tabular}    
\end{tabular}%
}
\end{table}

%% file: Sections/3-RelatedWork.tex
\section{Related Work} \label{sec:related-work}
This section reviews three bodies of literature that frame \ac{zt}-\ac{cps} convergence: domain-level convergence studies, technical adaptations for constrained and legacy environments, and risk- or safety-aware authorization.

\textbf{\ac{zt} Convergence with \ac{iot} and \acp{cps}}    \label{subsec:rw-convergence}
Existing surveys organize the \ac{zta} design space around architectural components, identity and access management, continuous authentication, contextual trust evaluation, and dynamic policy enforcement~\cite{Critical_ZT,ComprehensiveReview_ZT,ReviewComparative_ZT,Survey_ZT,Theory_ZT}.
They also identify persistent challenges involving conceptual ambiguity, policy complexity, verification overhead, and migration cost~\cite{Critical_ZT}.
These works establish the principles and implementation concerns of \ac{zt}, but do not systematically examine how its assumptions change when protected resources participate in physical processes that must remain observable and controllable.
Broader reviews treat \ac{iot}, industrial systems, healthcare, mobile networks, and critical infrastructures as application domains of \ac{zt}~\cite{SystematicReview_ZT, Verify_and_trust, Converging_ZT, ZT_ContextIoT}.
They recognize resource constraints, long lifecycles, legacy protocols, dynamic inventories, and incomplete device-state visibility as recurring deployment barriers~\cite{ALABA2017, HASSIJA2019, Ratasich, Tange, SystematicReview_ZT, Converging_ZT, ZT_ContextIoT}.
Long-lived and constrained devices can make continuous reassessment unsustainable, whereas decentralized or mobile populations challenge the completeness and freshness of centralized inventories and policy state. 
Dedicated convergence studies respond through two main adaptation directions.
The first strengthens devices and protocols through authentication, protected communication, credential management, and continuous monitoring.
The second accommodates constrained or non-modifiable assets through gateways, segmentation, monitoring, and external trust or policy services~\cite{Verify_and_trust,Converging_ZT,ZT_ContextIoT}.

This literature explains why \ac{iot} conditions make continuous assessment and granular enforcement difficult, but generally treats the resulting problems as a common set of implementation barriers.
It rarely distinguishes difficulties inherent to \ac{zt}, challenges amplified by \ac{iot} operating conditions, and strains arising specifically from cyber-physical coupling.
\textit{Wehbe et al.}~\cite{Converging_ZT}, for example, explicitly compare \ac{zt} expectations with \ac{iot} limitations, but do not identify the physical consequences of enforcement as a separate class of \ac{cps}-specific strain.

\textbf{Technical Adaptations for (I)\ac{iot} and \acp{cps}}
Technical proposals address three recurring needs: obtaining reliable device evidence, accommodating constrained or non-modifiable assets, and scaling policy enforcement~\cite{SystematicReview_ZT, Verify_and_trust, TrustAware_ZT,ZT_ContextIoT}.
Evidence-oriented mechanisms strengthen authentication, posture assessment, and continuous authorization, improving the basis for determining whether an entity or request should be trusted.
Stronger evidence can establish identity or integrity, but not whether an authenticated measurement is physically correct or whether removing it leaves sufficient information for operation.
These mechanisms also generally evaluate the requesting component or communication without distinguishing the physical roles of observation and actuation.
Revoking command authority may therefore remain coupled to suppressing read-only telemetry.

A second family of approaches uses gateways, SDN/NFV, micro-segmentation, and remote-access controls to enforce policies on behalf of assets that cannot host local \ac{zt} mechanisms~\cite{Converging_ZT,ZT_ContextIoT,SystematicReview_ZT}.
These techniques extend the available enforcement surface, but their primitives remain primarily cyber-operational, namely they allow, block, route, isolate, segment, or mediate network flows.
Network placement and flow-level attributes do not by themselves reveal whether traffic carries telemetry, estimator inputs, operator feedback, or actuator commands.
Consequently, these mechanisms cannot alone differentiate enforcement according to the physical function and consequence of each communication path.

\textbf{Risk- and Safety-Aware \acp{cps}}
Research on \acp{cps} moves beyond deployment feasibility by modeling how malicious or unreliable actions affect the physical process.
More advanced approaches incorporate this impact into trust evaluation or policy selection.
Recent examples include risk-aware protection of power-grid supply chains against generative-AI threats~\cite{MUNIR2024}, safety-relevant \ac{zt} adoption in connected vehicles~\cite{Towards_ZT}, and impact-aware runtime enforcement for foundation-model-controlled industrial robots~\cite{RANATHUNGA2026}. 
These works make physical consequence relevant to security decisions, but focus primarily on harm caused by malicious information or delegated actuation rather than harm introduced by the defensive response itself.

The Cyber-Physical \ac{zta} (CP-ZTA) proposed by \textit{Feng et al.}~\cite{CP-ZTA} is the closest to the decision problem considered here.
It combines a multi-layer access-control engine with a physical-model-based and data-driven policy optimizer, allowing physical-process information to influence trust evaluation and policy selection. However, this information is used primarily to estimate compromise risk and select restrictions on communication and lateral movement.
It does not separately model restriction-induced consequences, such as loss of operationally relevant telemetry, observability, or operator awareness, nor does it define graduated states that independently regulate measurement visibility, automated influence, and control authority.\\

Overall, two related gaps remain in the existing literature.
Analytically, prior work does not systematically distinguish general \ac{zt} difficulties from strains amplified by \ac{iot} or introduced by cyber-physical coupling.
Operationally, it offers limited guidance on which capabilities should be withdrawn once a component is judged risky, or whether the chosen restriction creates a separate physical consequence.
Addressing these gaps requires a cause-aware analysis of convergence strains and an enforcement model that keeps residual cyber risk and restriction-induced physical consequences distinct while separating observation, automated influence, and state-changing authority.

%% file: Sections/4-ZT_strains.tex
\section{ZT Convergence Strains and Operational Requirements in IoT/CPSs}   \label{sec:zt-iot-strains}
\ac{iot} and \acp{cps} alter the actors, evidence, communication patterns, and enforcement surfaces on which \ac{zt} decisions depend.
We represent these effects through a three-stage relation.
A \emph{domain property} is an inherent deployment characteristic, such as constrained endpoints, headless operation, or dependence on timely sensor data.
A \emph{convergence strain} (S1--S9) is the mismatch that this property creates with one or more \ac{zt} tenets.
An \emph{operational requirement} (R1--R9) specifies the capability needed to address that mismatch.

We classify a strain as \emph{IoT-amplified} when it has an enterprise-IT analogue but becomes systematic or materially harder under \ac{iot} conditions.
In contrast, a strain is \emph{CPS-specific} when its distinctive significance arises from coupling communication and computation with a physical process and cannot be removed merely by increasing endpoint resources or scaling conventional policy infrastructure.
Cross-cutting concerns such as privacy, explainability, governance, policy conflicts, and control-plane security remain relevant but are not treated as separate strains because they also apply to enterprise \ac{zt}~\cite{Survey_ZT,Critical_ZT,ReviewComparative_ZT,ComprehensiveReview_ZT,SystematicReview_ZT}. 
Table~\ref{tab:zt-iot-strains} records each strain's affected tenets, origin, and corresponding requirement.

\input{Tables/ZT-strain}

\subsection{IoT-Amplified Strains}  \label{subsec:iot-amplified-strains}
These strains also occur in enterprise IT, but \ac{iot} increases their scale, persistence, or implementation difficulty.

\textbf{S1 -- Resource asymmetry.}
Communication protection, posture assessment, dynamic authentication, and detailed telemetry can exceed the computational or energy budgets of constrained devices~\cite{Survey_ZT, ZT_ContextIoT, Converging_ZT, Verify_and_trust}.
Gateways can offload analysis and selected security operations, but cannot fully replace endpoint evidence about software integrity and internal state.
Security tasks may therefore compete with sensing, control, and other primary functions~\cite{HASSIJA2019,HaddadPajouh}.
Uniform assurance mechanisms and reassessment rates are consequently unsustainable across heterogeneous endpoints.
\emph{Capability-proportionate continuous assurance (R1)} requires evidence collection and analysis to match endpoint capabilities, for example through compact evidence, event-triggered attestation, or offloaded analysis~\cite{Tange,ZT_ContextIoT,Converging_ZT}.
The attainable evidence quality, maximum reassessment interval, and residual risk between assessments must remain explicit and auditable.

\textbf{S2 -- Autonomous and headless machine operation.}
Enterprise access control commonly assumes an interactive user, managed workstation, and bounded session~\cite{Theory_ZT,ReviewComparative_ZT}.
Many \ac{iot} devices instead operate autonomously, weakening assumptions about consent, multifactor authentication, and session termination~\cite{ALABA2017,HASSIJA2019,ZT_ContextIoT,Verify_and_trust,Towards_ZT}.
A device credential identifies the requester but does not establish approved firmware, current process role, location, or authority to perform a particular action~\cite{ZT_ContextIoT,TrustAware_ZT}.
\emph{Machine-centric identity and operational-authority governance (R2)} must therefore bind identity to operational role, process location, requested action, and machine-defined interaction boundaries, and revise authority when these attributes change.

\textbf{S3 -- Lifecycle mismatch and cryptographic aging.}
\ac{iot} combines rapidly replaced consumer devices with industrial controllers that may remain deployed for decades~\cite{ZT_ContextIoT,Converging_ZT,Ratasich,Tange}.
Vendor dependence, heterogeneous update mechanisms, and uneven support further complicate lifecycle governance~\cite{Ratasich,HASSIJA2019,HaddadPajouh}.
Identity, posture, and cryptographic state must nevertheless remain accurate and renewable even when certificates expire, firmware becomes unsupported, or a manufacturer disappears.
\emph{Lifecycle-resilient trust and cryptographic agility (R3)} requires enrollment, renewal, revocation, ownership transfer, maintenance, and decommissioning to update identity, posture, and authority consistently.
Trust must be recoverable after legitimate servicing and withdrawable when provenance or support can no longer be established, while algorithms and keys must remain replaceable without assuming simultaneous fleet-wide upgrades.

\textbf{S4 -- Fleet scale, mobility, and heterogeneity.}
Large fleets change connectivity, location, topology, ownership, and administrative control.
Devices may sleep, reconnect through different gateways, or cross network and administrative domains~\cite{ZT_ContextIoT,Converging_ZT,Verify_and_trust,FutureIndustry_ZT,Towards_ZT}.
Fleet scale and cyber-physical diversity also increase the cost of context analysis and per-component policy optimization~\cite{CP-ZTA}.
Centralized decision points simplify policy management but can become bottlenecks and availability dependencies.
On the other side, distributed decisions, edge enforcement, and cached policies introduce replication, consistency, and invalidation problems~\cite{Converging_ZT,CP-ZTA,SystematicReview_ZT}.
\emph{Scalable distributed visibility and policy coordination (R4)} therefore requires identity, posture, role, and policy state to remain aligned across central and local components.
Cached authority needs bounded validity and explicit invalidation, while policy distribution must define consistency, failover, and reconciliation guarantees~\cite{Converging_ZT,SystematicReview_ZT,ComprehensiveReview_ZT}.

\textbf{S5 -- Protocol and enforcement-surface heterogeneity.}
Devices expose capabilities through heterogeneous protocols, proprietary encodings, data models, and vendor-specific interfaces~\cite{ALABA2017,HASSIJA2019,Humayed}.
Telemetry, configuration, maintenance, software updates, and actuation may also share a device, session, or channel despite having different physical consequences~\cite{Humayed,Giraldo,Ding,Dibaji}.
Gateways that regulate only identities, addresses, ports, or flows cannot distinguish the operational authority carried within those communications~\cite{ZT_ContextIoT,Verify_and_trust,Converging_ZT}.
Effective least privilege enforcement therefore requires mapping vendor-specific operations to common semantic capabilities and enforcing policy at the same granularity.
\emph{Semantic-aware, operation-level least privilege (R5)} distinguishes observation, configuration, maintenance, software modification, and actuation across heterogeneous devices and protocols~\cite{Humayed,Ding,ZT_ContextIoT}.
It can thereby withdraw command or write authority while retaining constrained telemetry for diagnosis or awareness.

\textbf{S6 -- Legacy and non-modifiable assets.}
Many deployed \ac{iot} and industrial assets cannot host a \ac{zt} agent, expose a modern identity interface, or be modified without disrupting operation.
Gateways, proxies, and protocol mediators consequently provide external evidence collection and enforcement~\cite{ZT_ContextIoT,Converging_ZT,SystematicReview_ZT}.
Mediation, however, makes evidence indirect and enforcement coarser.
A gateway may secure upstream traffic while remaining blind to downstream posture, resource boundaries, or proprietary operations.
It may also concentrate compromise and bypass risk or recreate perimeter-based trust~\cite{ZT_ContextIoT,Converging_ZT}.
\emph{Legacy-compatible mediated enforcement (R6)} should preserve, where technically possible, downstream identity, resource-level policy, east-west isolation, and operation-level control.
Its assurance case must state evidence blind spots, trust assumptions, bypass resistance, integrity, availability, and policy-version consistency.
Where fine-grained enforcement is impossible, the remaining coarse authority must be recorded as residual risk rather than treated as equivalent to endpoint-native control.

\subsection{CPS-Specific Strains}   \label{subsec:cps-specific-strains}
\acp{cps} add physical-process semantics, operational deadlines, cross-layer effects, and the need to preserve critical functions during disruption~\cite{CP-ZTA,Towards_ZT,Ratasich,Tange,Humayed,Segovia-Ferreira}.
These conditions require authorization to be timely, physically meaningful, and aware of the consequences introduced by enforcement, leading to S7--S9.

\textbf{S7 -- Real-time operation and degraded connectivity.}
\ac{cps} operations may impose strict sampling periods, ordering constraints, and bounded delay or jitter~\cite{Ratasich,Tange,Towards_ZT}.
Remote identity providers, \acp{pdp}, revocation services, and telemetry pipelines can add variable latency or become unavailable, yet authorization must reach the \ac{pep} before the relevant operational deadline. 
Unlike an enterprise delay that primarily impairs service availability, a missed \ac{cps} deadline can invalidate a control action and change the physical trajectory.

Dynamic authorization therefore cannot always wait for complete global evidence.
Caching reduces latency only while the identity, posture, policy, and physical-state assumptions supporting the cached decision remain valid~\cite{Tange,Ratasich,Segovia-Ferreira}.
\emph{Real-time and disruption-tolerant authorization (R7)} requires latency bounds tied to operation criticality and local decision paths or pre-authorized contingencies for time-critical actions.
Cached authority and degraded operation must identify the responsible component, admissible evidence, scope, duration, state and temporal validity, and conditions for restriction, revocation, or recovery.
When a preferred response cannot be validated or installed in time, the system must invoke a declared fallback with explicit assumptions and recovery behavior, rather than assume that the most restrictive action is inherently safer~\cite{Ratasich,Kim,Segovia-Ferreira}.

\textbf{S8 -- Evidence uncertainty and cyber-physical semantics.}
In \acp{cps}, sensors, estimators, and controllers jointly generate evidence for authorization~\cite{Theory_ZT,TrustAware_ZT,CP-ZTA}.
Cryptographic protection can establish message origin and integrity, but not the physical correctness or operational legitimacy of the reported value.
Evidence must also be fresh, sufficiently independent, physically plausible, and relevant to the requested operation.
An authorization decision may therefore rely on authentic evidence and still be operationally invalid.

A single deterministic trust score can further obscure the distinction between evidence of low trust and insufficient evidence~\cite{Ding,Dibaji,CP-ZTA}.
\emph{Uncertainty-aware cyber-physical trust evaluation (R8)} requires decision context to represent provenance, freshness, physical plausibility, correlation, redundancy, and model confidence.
Uncertainty must influence authorization explicitly, and the selected action must remain explainable from the evidence and policy that produced it.
Because this context may expose device usage, location, production state, or human behavior, it must also be minimized, protected, and limited to its authorized purpose~\cite{Survey_ZT,Verify_and_trust,Converging_ZT}.

\textbf{S9 -- Response-induced physical risk and enforcement.}
Security and safety controls can reinforce or conflict with each other, and reducing cyber exposure does not necessarily reduce total cyber-physical risk~\cite{Kriaa}.
Removing a compromised component may contain an attack while also suppressing measurements needed for state estimation, disconnecting a controller needed for stabilization, or eliminating a recovery path required for intervention~\cite{Segovia-Ferreira}.
Unlike conventional IT denial, \ac{cps} enforcement changes the observation and control paths governing the physical process and may therefore introduce a separate physical consequence.

Least privilege must consequently be applied at capability level, and low trust need not imply total disconnection.
Instead, command or write authority may be withdrawn while selected telemetry remains visible for corroboration, diagnosis, or recovery.
Visibility does not, however, imply correctness, estimator admission, or automated influence.
These properties must be controlled separately and interpreted under the uncertainty represented by R8~\cite{Kriaa,Segovia-Ferreira}.
\emph{Physical-consequence-aware graduated enforcement and recovery (R9)} requires modes between unrestricted operation and complete isolation, including read-only operation, constrained actuation or telemetry, selective isolation, and a declared degraded or fallback state whose validity conditions are checked.
Candidate responses that violate declared hard admissibility conditions must be excluded rather than compensated for by favorable risk scores.
Among the remaining candidates, the decision process keeps residual cyber risk and restriction-induced physical consequences distinct, while recovery governs the staged restoration of authority, observability, and normal control~\cite{Dibaji,Segovia-Ferreira}.
R9 therefore requires the physical validity of a restriction to be evaluated rather than inferred from its severity.

Collectively, R1--R9 form three dependent layers.
R1--R6 establish whether \ac{zt} controls can be implemented across constrained, heterogeneous, dynamic, and non-modifiable environments.
R7--R8 require decisions to meet operational deadlines and rely on physically meaningful, uncertainty-aware evidence.
R9 adds consequence-aware graduated enforcement and recovery.

%% file: Tables/ZT-strain.tex
\begin{table*}[]
\renewcommand{\arraystretch}{1.25}
\centering
\caption{Map of ZT-IoT/CPS convergence strains to affected NIST ZT tenets, domain origin, and corresponding operational requirements.}
\label{tab:zt-iot-strains}
\resizebox{.9\textwidth}{!}{%
\begin{tabular}{c|c|c|c|c}
\rowcolor[HTML]{C0C0C0} 
\textbf{Strain ID} &
  \textbf{\begin{tabular}[c]{@{}c@{}}ZT-IoT/CPS\\ convergence strain\end{tabular}} &
  \textbf{\begin{tabular}[c]{@{}c@{}}ZT tenets under\\ strain\end{tabular}} &
  \textbf{\begin{tabular}[c]{@{}c@{}}Domain\\ origin\end{tabular}} &
  \textbf{\begin{tabular}[c]{@{}c@{}}Operational\\ Requirement\end{tabular}} \\ \hline
\textit{S1} &
  Resource asymmetry &
  T2, T5, T6, T7 &
  IoT-amplified &
  \begin{tabular}[c]{@{}c@{}}Capability-proportionate\\ continuous assurance\end{tabular} \\
\rowcolor[HTML]{EFEFEF} 
\textit{S2} &
  \begin{tabular}[c]{@{}c@{}}Autonomous and headless\\ machine operation\end{tabular} &
  T3, T4, T6 &
  IoT-amplified &
  \begin{tabular}[c]{@{}c@{}}Machine-centric identity and\\ operational-authority governance\end{tabular} \\
\textit{S3} &
  \begin{tabular}[c]{@{}c@{}}Lifecycle mismatch and\\ cryptographic aging\end{tabular} &
  T2, T4, T5, T6 &
  \begin{tabular}[c]{@{}c@{}}IoT-amplified;\\ especially severe in CPS\end{tabular} &
  \begin{tabular}[c]{@{}c@{}}Lifecycle-resilient trust and\\ cryptographic agility\end{tabular} \\
\rowcolor[HTML]{EFEFEF} 
\textit{S4} &
  \begin{tabular}[c]{@{}c@{}}Fleet scale, mobility, and\\ administrative heterogeneity\end{tabular} &
  T1, T4, T5, T6, T7 &
  IoT-amplified &
  \begin{tabular}[c]{@{}c@{}}Scalable distributed visibility and\\ policy coordination\end{tabular} \\
\textit{S5} &
  \begin{tabular}[c]{@{}c@{}}Protocol and enforcement-surface\\ heterogeneity\end{tabular} &
  T2, T3, T4, T6 &
  IoT-amplified &
  \begin{tabular}[c]{@{}c@{}}Semantic-aware, operation-level\\ least privilege\end{tabular} \\
\rowcolor[HTML]{EFEFEF} 
\textit{S6} &
  \begin{tabular}[c]{@{}c@{}}Brownfield and\\ non-modifiable assets\end{tabular} &
  T2, T3, T5, T6, T7 &
  IoT-amplified/CPS &
  \begin{tabular}[c]{@{}c@{}}Brownfield/legacy-compatible\\ mediated enforcement\end{tabular} \\
\textit{S7} &
  \begin{tabular}[c]{@{}c@{}}Real-time operation and\\ degraded connectivity\end{tabular} &
  T3, T4, T6, T7 &
  \begin{tabular}[c]{@{}c@{}}CPS-specific\\ in consequence\end{tabular} &
  \begin{tabular}[c]{@{}c@{}}Real-time and disruption-tolerant\\ authorization\end{tabular} \\
\rowcolor[HTML]{EFEFEF} 
\textit{S8} &
  \begin{tabular}[c]{@{}c@{}}Evidence uncertainty and\\ cyber--physical semantics\end{tabular} &
  T4, T5, T6, T7 &
  \begin{tabular}[c]{@{}c@{}}IoT-amplified for evidence;\\ CPS-specific for semantics\end{tabular} &
  \begin{tabular}[c]{@{}c@{}}Uncertainty-aware cyber-physical\\ trust evaluation\end{tabular} \\
\textit{S9} &
  \begin{tabular}[c]{@{}c@{}}Response-induced physical risk\\ and enforcement\end{tabular} &
  T3, T4, T6, T7 &
  CPS-specific &
  \begin{tabular}[c]{@{}c@{}}Physical-safety-preserving graduated\\ enforcement and recover\end{tabular}
\end{tabular}%
}
\end{table*}

%% file: Sections/5-SafetyAware_ZT.tex
\section{Safety-Aware Zero Trust} \label{sec:general-safety-engine}
Conventional trust evaluation determines whether available security evidence supports authorization, but not which restriction should follow when several responses are possible and each may affect the physical process differently.
We address this gap proposing \acrfull{sa-zt}, a framework that extends \ac{nist} \ac{zta} with consequence-aware response selection and capability-level enforcement.
This section defines its architecture, principles, enforcement semantics, and selection rule.
Sec.~\ref{sec:ieee30-safety-engine-instantiation} provides an illustrative IEEE 30-bus instantiation.

\subsection{Architecture} \label{subsec:general-reference-architecture}
\ac{sa-zt} separates the \emph{command path}, which carries state-changing operations such as writes, configuration changes, and actuation, from the \emph{telemetry path}, which carries measurements used by operators, estimators, controllers, and protection functions.
It augments the \ac{nist} \ac{zta} model with two logical components: a \textbf{\acrfull{se}} associated with the \ac{pe}, and a \textbf{\acrfull{tb}} acting as a specialized \ac{pep} on the telemetry path.
The \ac{se} receives the conventional trust output and current operating context, eliminates responses that violate declared hard conditions, and ranks the remainder by residual cyber risk and the consequences of enforcement.
The \ac{pa} compiles the selected configuration into versioned rules for command-path \acp{pep}, the \ac{tb}, and relevant local controllers.
The \ac{tb} then enforces which telemetry remains visible and the maximum influence it can have on automated functions, allowing a suspicious measurement, for example, to remain available for operator awareness while being excluded from an estimator.
Every telemetry path capable of affecting an automated decision must therefore traverse a trusted broker or equivalent mediation point.

The \ac{se} and \ac{tb} are logical roles, not necessarily distinct physical nodes.
They may be centralized, distributed across process zones, or organized hierarchically, provided that, as required by R4, coupled decisions are coordinated or assigned to explicit, non-overlapping scopes.
Distributed \acp{pep} must enforce the same configuration version within their scopes and report installation status.
If a coordinated decision cannot be installed within its deadline, the affected component invokes the fallback defined under R7, with explicit assumptions, scope, validity conditions, duration, locally retained authority, and recovery procedure.

\ac{sa-zt} operates as a closed decision cycle, meaning that security and process evidence update the cyber belief and operating context.
Thus, the \ac{se} selects an admissible response, the \ac{pa} distributes it, and the enforcement components apply its authority, visibility, influence, and degraded-mode semantics.
The resulting observations inform the next decision epoch.
For auditability, each decision record groups the evidence and uncertainty used, evaluated candidates and objective terms, binding hard conditions, selected modes and configuration version, and any installation failure or fallback.

\subsection{Decision Context and Capability-Level Enforcement} \label{subsec:general-decision-context}
Table~\ref{tab:general-notation} in Appendix~\ref{sec:Terminology} summarizes the notation.

\textbf{Decision units and candidate configurations.}
At decision epoch \(k\), the \ac{se} assigns enforcement modes to the independently controllable units in \(\mathcal{E}=\{1,\ldots,N\}\), where \(N=|\mathcal{E}|\).
A unit is the smallest scope that can be enforced independently, such as a device, logical role, or resource-operation pair.
For each unit \(i\), \(\Sigma_i\) is its supported mode set, and \(\sigma_{i,k}\in\Sigma_i\) is the mode assigned at epoch \(k\).
The joint configuration \(\boldsymbol{\sigma}_k=(\sigma_{1,k},\ldots,\sigma_{N,k})\) therefore specifies the response across the complete enforcement scope.
Representing independently mediated capabilities as separate units enables observation and actuation to be restricted differently, as required by R5.

Candidates are evaluated from two complementary state descriptions.
The \emph{cyber belief} \(\mathcal{B}_k\) represents the compromises and attack paths supported by the available security evidence, whereas the \emph{operating context} \(\boldsymbol{\xi}_k\) represents the estimated process state, operating mode, objectives, deadlines, and uncertainty.
Both contain only information available at epoch \(k\).
Predictions may use current evidence and declared forecasts, but not subsequently realized disturbances, delays, or attack outcomes.
An estimator output may enter \(\boldsymbol{\xi}_k\) only under its stated validity conditions.

The candidate space is filtered in two stages.
First, \(\mathcal{A}_k\) contains configurations that are currently authorized, supported, and installable given the available \acp{pep}, endpoint modes, response deadlines, resource budgets, dwell-time constraints, and recovery state.
Second, \(\mathcal{F}_k\subseteq\mathcal{A}_k\) retains only configurations satisfying every declared hard operational and physical condition.
Additional filtering applies to the joint configuration, where individually acceptable restrictions may together remove model-based observability, required feedback, or a recovery path, while individually limited capabilities may together preserve an attack path.
Soft objective terms rank only candidates in \(\mathcal{F}_k\) and cannot compensate for a hard violation.

\textbf{Capability-decoupling interface.}
\ac{sa-zt} expresses the enforcement effect of a configuration as
\begin{equation}
\mathcal{D}(\boldsymbol{\sigma})=
\left(
\mathcal{U}(\boldsymbol{\sigma}),
\mathcal{V}(\boldsymbol{\sigma}),
\mathcal{W}(\boldsymbol{\sigma})
\right),
\label{eq:general-decoupling}
\end{equation}
where \(\mathcal{U}(\boldsymbol{\sigma})\) is the retained state-changing authority; \(\mathcal{V}(\boldsymbol{\sigma})\) is the raw telemetry retained and made visible at authorized ingress points; and \(\mathcal{W}(\boldsymbol{\sigma})\) is the maximum contribution that this telemetry may make to estimates, control decisions, protection functions, or other automated consumers.
Raw visibility does not imply correctness, trust, estimator admission, or automated influence.
Similarly, observability denotes structural sufficiency under the applicable system model, not merely the amount of visible data.

Separating \(\mathcal{V}(\boldsymbol{\sigma})\) from \(\mathcal{W}(\boldsymbol{\sigma})\) allows a suspicious observation to remain available for awareness or forensic analysis while its automated influence is reduced or removed.
It also prevents command blocking from being mistaken for complete containment: falsified telemetry may still alter the physical process through an estimator or controller~\cite{CP-ZTA}.
Applications may implement influence limits through admission rules, estimator weights, confidence bounds, routing, or restrictions on eligible consumers.
At runtime, the \ac{tb} applies the uncertainty-aware treatment required by R8 and may refine \(\mathcal{W}(\boldsymbol{\sigma})\) only within a prevalidated refinement family or after reevaluating the admissibility of the affected estimator and plant configuration.
If that validity cannot be established, it invokes the declared fallback rather than assuming that further restriction is safer.

\textbf{Graduated enforcement modes.}
Eq.~\eqref{eq:general-decoupling} defines a response space between unrestricted operation and complete isolation.
\ac{sa-zt} uses five reference intents: \textsf{Full} for policy-authorized normal operation; \textsf{Guarded} for bounded least-privilege continuation; \textsf{Observation-only} for monitoring without state-changing authority; \textsf{Safe-local} for predefined local fallback or protection functions; and \textsf{Isolated} for strong containment, possibly with a separately protected health, re-attestation, or recovery path.
These are enforcement intents, not universally ordered safety or restrictiveness levels.
Their concrete semantics and supported subsets are application-specific and are summarized in Table~\ref{tab:general-enforcement-modes} through \(\mathcal{U}(\boldsymbol{\sigma})\), \(\mathcal{V}(\boldsymbol{\sigma})\), and \(\mathcal{W}(\boldsymbol{\sigma})\).
Restoring broader authority is a new authorization decision and may require renewed attestation, functional validation, minimum dwell time, and staged recovery.

\input{Tables/EnforcementModes}

\subsection{Safety-Engine Formulation} \label{subsec:general-safety-rule}
For each admissible configuration, the \ac{se} evaluates four nonnegative terms: residual adversarial risk \(R_{\mathrm{cyb},k}\), non-safety-critical operational degradation \(C_{\mathrm{op},k}\), response-induced physical risk \(R_{\mathrm{resp},k}\), and non-physical switching or recovery cost \(C_{\mathrm{sw},k}\).
It selects
\begin{equation}
\begin{aligned}
\boldsymbol{\sigma}_k^\star
\in \arg\min_{\boldsymbol{\sigma}\in\mathcal{F}_k}
\Big[
&\lambda_{\mathrm{cyb}} R_{\mathrm{cyb},k}
(\boldsymbol{\sigma};\mathcal{B}_k,\boldsymbol{\xi}_k) \\
&+\lambda_{\mathrm{op}} C_{\mathrm{op},k}
(\boldsymbol{\sigma};\boldsymbol{\xi}_k) \\
&+\lambda_{\mathrm{resp}} R_{\mathrm{resp},k}
(\boldsymbol{\sigma};\mathcal{B}_k,\boldsymbol{\xi}_k) \\
&+\lambda_{\mathrm{sw}} C_{\mathrm{sw},k}
(\boldsymbol{\sigma},\boldsymbol{\sigma}_{k-1})
\Big].
\end{aligned}
\label{eq:general-safety-engine}
\end{equation}
The nonnegative coefficients encode policy priorities and regulate trade-offs only within \(\mathcal{F}_k\).
Residual cyber risk captures the threat retained with the allowed capabilities; the other terms capture consequences introduced by restricting or reconfiguring them.
The optimization therefore selects a response to the current cyber belief and operating context rather than reassessing endpoint trust.

Eq.~\eqref{eq:general-safety-engine} defines the framework-level objective, but it does not require every instantiation to solve the complete joint problem globally.
Thus, an implementation may use a causally staged or bounded procedure if it preserves the meaning of the terms, enforces the declared hard conditions, respects information timing, and reports its search scope.
The IEEE 30-bus instantiation in Sec.~\ref{sec:ieee30-safety-engine-instantiation} follows this approach and does not claim global optimality.
Hard admissibility remains separate from ranking: no objective value can admit a configuration outside \(\mathcal{F}_k\).
Accordingly, \(\lambda_{\mathrm{cyb}}>0\), while \(\lambda_{\mathrm{resp}}>0\) is required unless all relevant response-induced hazards are already hard conditions.
Ties may be resolved deterministically by lower response-induced risk, lower residual cyber risk, and then fewer changes from \(\boldsymbol{\sigma}_{k-1}\).
If \(\mathcal{F}_k=\varnothing\), the engine reports the violated conditions and invokes the declared degraded-mode fallback instead of selecting an infeasible configuration.

\textbf{Residual cyber risk.}
\(R_{\mathrm{cyb},k}(\boldsymbol{\sigma};\mathcal{B}_k,\boldsymbol{\xi}_k)\) measures the adversarial capability or expected consequence remaining under \(\mathcal{D}(\boldsymbol{\sigma})\).
Depending on the application, it may use endpoint exposure, attack-graph reachability, joint attack feasibility, or consequence conditioned on exploitation.
Physical harm caused by a successful attack belongs to this term, not to response-induced risk.

\textbf{Operational degradation.}
\(C_{\mathrm{op},k}(\boldsymbol{\sigma};\boldsymbol{\xi}_k)\) captures legitimate-performance loss caused by restriction, such as reduced estimation quality, control flexibility, service continuity, or increased operator workload.
It excludes consequences classified by the application as physical harm.

\textbf{Response-induced physical risk.}
\(R_{\mathrm{resp},k}(\boldsymbol{\sigma};\mathcal{B}_k,\boldsymbol{\xi}_k)\) captures physical consequences attributable to enforcing \(\boldsymbol{\sigma}\), distinct from attack-induced harm.
Let \(C_{\mathrm{phys},k}^{\boldsymbol{\sigma},0}\) be the predicted physical consequence over a declared horizon under the candidate response and \(C_{\mathrm{phys},k}^{\mathrm{ref},0}\) the consequence under a fixed reference response.
Define
\begin{equation}
\Delta C_{\mathrm{phys},k}^{\boldsymbol{\sigma},0} =
C_{\mathrm{phys},k}^{\boldsymbol{\sigma},0}
-C_{\mathrm{phys},k}^{\mathrm{ref},0}.
\label{eq:general-response-difference}
\end{equation}
Positive values indicate additional predicted harm, zero no change, and negative values a lower predicted consequence.
The superscript \(0\) means that no additional adversarial action is introduced after epoch \(k\).
Prior attack effects remain represented in \(\mathcal{B}_k\) and \(\boldsymbol{\xi}_k\).
With the notation \((a)_+=\max\{a,0\}\),
\begin{equation}
\begin{aligned}
R_{\mathrm{resp},k}
(\boldsymbol{\sigma};\mathcal{B}_k,\boldsymbol{\xi}_k)
={}&
\mathbb{E}\!\left[
\left(\Delta C_{\mathrm{phys},k}^{\boldsymbol{\sigma},0}\right)_{+}
\,\middle|\,
\mathcal{B}_k,\boldsymbol{\xi}_k
\right],
\end{aligned}
\label{eq:general-response-risk}
\end{equation}
where \(\mathbb{E}[\cdot]\) is the expected value.
The reference may be \(\boldsymbol{\sigma}_{k-1}\), continuation without further restriction, or another declared baseline, but must remain fixed across candidates at epoch \(k\).
Eq.~\eqref{eq:general-response-risk} retains only incremental harm attributable to enforcement, while containment benefit remains in \(R_{\mathrm{cyb},k}\).
If no probabilistic predictor is available, the expectation may be replaced by a declared conservative bound or other risk measure based only on information available at epoch \(k\).
Any material attack-response interaction must be represented once rather than duplicated across the two risk terms.

\textbf{Switching and recovery cost.}
\(C_{\mathrm{sw},k}(\boldsymbol{\sigma},\boldsymbol{\sigma}_{k-1})\) penalizes unnecessary mode changes, oscillation, reconfiguration overhead, and premature restoration.
Physical transition consequences belong to \(R_{\mathrm{resp},k}\); \(C_{\mathrm{sw},k}\) contains only non-physical costs.

Eq.~\eqref{eq:general-safety-engine} provides a consequence-aware selection rule; it does not prove universal safety or control-performance improvement.
Its claims are bounded by the validity, uncertainty, prediction horizon, and operating envelope of the models used to define \(\mathcal{F}_k\), \(R_{\mathrm{cyb},k}\), \(C_{\mathrm{op},k}\), and \(R_{\mathrm{resp},k}\).
The next section makes these assumptions explicit for the power-system instantiation.

%% file: Tables/EnforcementModes.tex
\begin{table*}[]
\centering
\caption{Capability semantics of the reference graduated enforcement modes.}
\label{tab:general-enforcement-modes}

\newcommand{\rowpad}{\rule[-6pt]{0pt}{26pt}}
\newcommand{\rowpadd}{\rule[-6pt]{0pt}{30pt}}

\resizebox{.9\textwidth}{!}{%
\begin{tabular}{l|c|c|c}
\hline
\rowcolor[HTML]{C0C0C0} 
\multicolumn{1}{c|}{\cellcolor[HTML]{C0C0C0}\textbf{Mode}} &
  \textbf{\begin{tabular}[c]{@{}c@{}}State-changing authority\\ \(\mathcal{U}(\sigma)\)\end{tabular}} &
  \textbf{\begin{tabular}[c]{@{}c@{}}Raw visibility\\ \(\mathcal{V}(\sigma)\)\end{tabular}} &
  \textbf{\begin{tabular}[c]{@{}c@{}}Maximum influence\\ \(\mathcal{W}(\sigma)\)\end{tabular}} \\
\hline

\textsf{Full}\rowpad &
  All operations authorized. &
  All telemetry available. &
  Ordinary policy. \\

\textsf{Guarded}\rowpad &
  \begin{tabular}[c]{@{}c@{}}Bounded by operation, resource,\\ value, rate, or validity interval.\end{tabular} &
  \begin{tabular}[c]{@{}c@{}}Selected or raw telemetry\\ remains available.\end{tabular} &
  \begin{tabular}[c]{@{}c@{}}Reduced or subject to\\ additional validation.\end{tabular} \\

\textsf{Observation-only}\rowpad &
  No remote state-changing operations. &
  \begin{tabular}[c]{@{}c@{}}Selected telemetry remains available\\ for monitoring and diagnosis.\end{tabular} &
  Disabled or limited \\

\textsf{Safe-local}\rowpad &
  \begin{tabular}[c]{@{}c@{}}Remote authority removed or constrained;\\ predefined local protection remains.\end{tabular} &
  \begin{tabular}[c]{@{}c@{}}Only reporting, validation, and recovery paths\\ required in degraded mode remain available.\end{tabular} &
  \begin{tabular}[c]{@{}c@{}}Limited to predefined local safety\\ and recovery functions.\end{tabular} \\

\textsf{Isolated}\rowpadd &
  \begin{tabular}[c]{@{}c@{}}No operational authority;\\ only separately protected recovery actions.\end{tabular} &
  \begin{tabular}[c]{@{}c@{}}No normal telemetry; only protected health,\\ recovery, or re-attestation paths, if declared.\end{tabular} &
  \begin{tabular}[c]{@{}c@{}}No influence;\\ only protected recovery behavior.\end{tabular} \\

\hline
\end{tabular}%
}
\end{table*}

%% file: Sections/6-IEEE30-Model.tex
\section{IEEE 30-Bus Power-Grid Safety-Engine Instantiation}    \label{sec:ieee30-safety-engine-instantiation}
This section instantiates \ac{sa-zt} on the IEEE 30-bus system distributed with MATPOWER~\cite{MATPOWER} and also considered by \textit{Feng et al.}~\cite{CP-ZTA}.
The case study maps the framework to a concrete plant, control architecture, enforcement interface, and set of Safety-Engine terms.
It is not proposed as a new grid controller or as evidence of certified safety; rather, it shows how capability-specific enforcement, estimator validity, physical admissibility, and authorization timing can be represented in an auditable decision process.
Scenario parameters and assessment horizons are introduced in Sec.~\ref{sec:ieee30-experimental-evaluation}, while Appendix~\ref{app:ieee30-formulation-details} provides the estimator, dispatch, enforcement, and dynamic-response details.

\subsection{System Model and Architectural Realization} \label{subsec:ieee30-system-model}
\begin{figure}[t]
    \centering
    \includegraphics[width=\columnwidth]{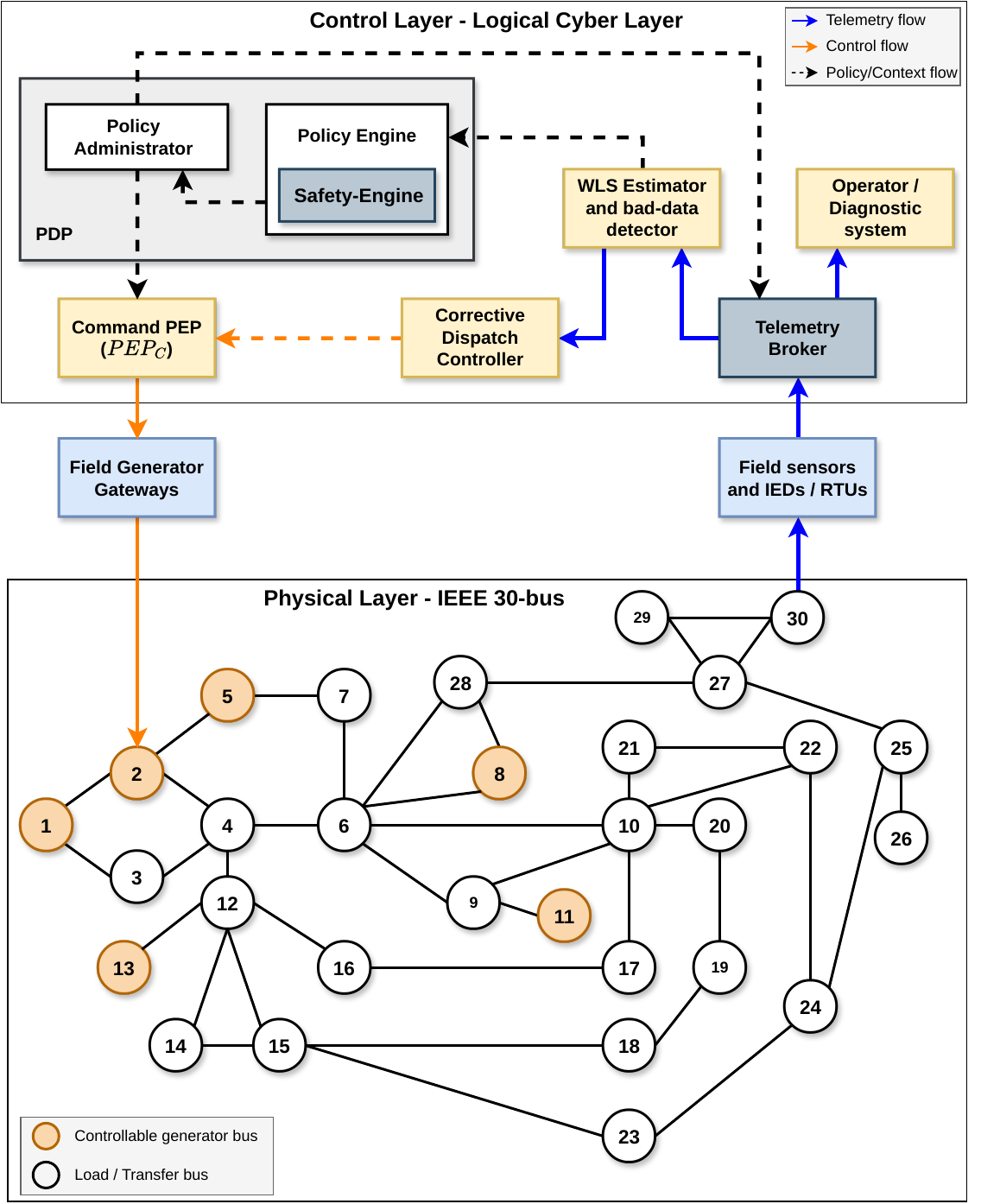}
    \caption{IEEE 30-bus \ac{sa-zt} instantiation. Orange buses host controlled generator interfaces.}
    \label{fig:ieee30-system-model}
\end{figure}

Fig.~\ref{fig:ieee30-system-model} shows the quasi-static direct-current (DC) active-power model, comprising 30 buses, 41 branches, and controllable generator interfaces at buses 1, 2, 5, 8, 11, and 13.
Bus~1 is the angular reference.
The field layer contains measurement transducers, substation \acp{ied}/\acp{rtu}, and generator controllers and actuators.
Branch-flow and bus-injection measurements enter the control center through the \ac{tb}, which acts as the telemetry-side \ac{pep}.
Dispatch setpoints leave through the command-side \ac{pep}, \(\mathrm{PEP}_{C}\).
The \ac{se} operates within the \ac{pe} of the \ac{pdp}.
The telemetry and command endpoints are cyber interfaces associated with plant elements, not additional electrical buses.

The implementation follows a causal two-stage decision order.
At epoch \(k\), the \ac{se} first selects telemetry modes from the defender-visible compromise belief and preceding configuration, before observing the sample governed by that decision.
After the \ac{tb} activates the selected telemetry configuration, the current sample is acquired and evaluated by the bad-data detector and \ac{wls} estimator.
Only after the estimate is accepted, or a declared fallback estimate is selected, the \ac{se} chooses generator-command modes.
The \ac{pa} distributes each decision as a consistent policy version, and the \(\mathrm{PEP}_{C}\) then applies the selected modes to the single dispatch request computed for that epoch.
Thus, telemetry selection cannot use the current sample, whereas command selection may use the accepted current estimate.

\textbf{Concrete enforcement semantics.}
On the telemetry path, \textsf{Full} preserves peer authority, raw visibility, and ordinary estimator precision.
\textsf{Guarded} reduces peer authority and the measurement's relative \ac{wls} precision while preserving the raw stream and without altering the physical sensor-noise distribution.
\textsf{Observation-only} retains the stream for explicitly authorized monitoring or diagnosis but removes its estimator and automatic-control influence, whereas \textsf{Isolated} removes ordinary visibility and automated use.
Telemetry endpoints do not support \textsf{Safe-local}.

On the command path, \textsf{Full} admits the dispatch request within the ordinary capacity and per-epoch flexibility envelope, while \textsf{Guarded} restricts the permitted change within that envelope.
In this quasi-static realization, \textsf{Safe-local} rejects the remote request and holds the previous validated output.
\textsf{Isolated} applies the same generation law but also removes the ordinary remote interface.
The modes therefore differ in authorization and recovery semantics, not in their modeled steady-state output.
Only \textsf{Full} and \textsf{Guarded} generators remain eligible for central balancing, and command endpoints do not support \textsf{Observation-only}.

\subsection{Mathematical Formulation}
\label{subsec:ieee30-compact-formulation}
Let \(\mathcal{N}=\{1,\ldots,30\}\) be the bus set, \(\mathcal{L}\subseteq\mathcal{N}\times\mathcal{N}\) the 41-branch set, and \(\mathcal{G}=\{1,2,5,8,11,13\}\) the generator-bus set.
The telemetry block contains one oriented measurement for each branch and one injection measurement at each of the 14 buses in \(\mathcal{N}^{p}\subseteq\mathcal{N}\).
Superscripts \(z\) and \(u\) identify telemetry-side and command-side quantities, respectively, while \(f\) and \(p\) distinguish branch-flow and bus-injection telemetry.
Accordingly, \(\mathcal{E}^{z}\) is the set of telemetry enforcement units and \(\mathcal{E}^{u}\) is the set of generator-command enforcement units. 
Thus,
\begin{equation}
    \begin{aligned}
    \mathcal{E}^{z}
        &\triangleq
          \{e^{f}_{\ell}:\ell\in\mathcal{L}\}
          \mathbin{\dot\cup}
          \{e^{p}_{n}:n\in\mathcal{N}^{p}\},
          &|\mathcal{E}^{z}|&=55,\\
    \mathcal{E}^{u}
        &\triangleq
          \{e^{u}_{g}:g\in\mathcal{G}\},
          &|\mathcal{E}^{u}|&=6,\\
    \mathcal{E}
        &=\mathcal{E}^{z}\mathbin{\dot\cup}\mathcal{E}^{u},
          &\boldsymbol{\sigma}_{k}
          &=\bigl(\boldsymbol{\sigma}^{z}_{k},
                   \boldsymbol{\sigma}^{u}_{k}\bigr).
    \end{aligned}
\label{eq:ieee30-enforcement-units}
\end{equation}
Here, \(e^{f}_{\ell}\), \(e^{p}_{n}\), and \(e^{u}_{g}\) denote concrete logical enforcement units associated with branch-flow measurement \(\ell\), injection measurement \(n\), and generator-command interface \(g\), respectively.
The telemetry vector \(\boldsymbol{\sigma}^{z}_k\) follows the 55 logical endpoint labels in Fig.~\ref{fig:ieee30-system-model}, and the command vector \(\boldsymbol{\sigma}^{u}_k\) follows the generator order in \(\mathcal{G}\).
The fixed row mapping is specified in Appendix~\ref{app:ieee30-formulation-details}.

\textbf{Plant, measurement, and estimation interface.}
With bus~1 fixed, \(\mathbf{x}_{k}\in\mathbb{R}^{29}\) contains the remaining voltage angles.
The matrix \(\mathbf{H}^{f}\in\mathbb{R}^{41\times29}\) maps this state to the oriented per-unit branch-flow measurements, whereas \(\mathbf{H}^{p}\in\mathbb{R}^{14\times29}\) maps it to the measured per-unit bus injections.
If \(\boldsymbol{\Pi}_{z}\) is the fixed permutation that places the stacked measurements in logical endpoint order, then
\begin{equation}
    \mathbf{H}
    =\boldsymbol{\Pi}_{z}
    \begin{bmatrix}
        \mathbf{H}^{f}\\
        \mathbf{H}^{p}
    \end{bmatrix}
    \in\mathbb{R}^{55\times29},
    \qquad
    \operatorname{rank}(\mathbf{H})=29.
    \label{eq:ieee30-H-construction}
\end{equation}
Appendix~\ref{app:ieee30-formulation-details} derives both \(\mathbf{H}^{f}, \mathbf{H}^{p}\).
Before policy-mediated estimation, the received telemetry is
\begin{equation}
    \mathbf{z}^{\mathrm{raw}}_{k}
    =\mathbf{H}\mathbf{x}_{k}
    +\boldsymbol{\eta}_{k}
    +\mathbf{a}_{k},
    \qquad
    \boldsymbol{\eta}_{k}\sim\mathcal{N}(\mathbf{0},\mathbf{R}_{k}),
\label{eq:ieee30-measurement-model}
\end{equation}
where \(\boldsymbol{\eta}_{k}\) is physical measurement noise with covariance \(\mathbf{R}_{k}\), and \(\mathbf{a}_{k}\) is the attack realized through compromised telemetry endpoints.
Policy-dependent precision is applied separately and is not part of \(\mathbf{R}_{k}\).

For telemetry configuration \(\boldsymbol{\sigma}^{z}\), the binary vector \(\mathbf{v}^{z}(\boldsymbol{\sigma}^{z})\in\{0,1\}^{55}\) records raw visibility.
Its \(j\)-th component is one exactly when telemetry stream \(j\) remains available to an authorized consumer.
The vector \(\boldsymbol{\omega}^{z}(\boldsymbol{\sigma}^{z})\in[0,1]^{55}\) records the relative \ac{wls} precision retained for each stream; \(\omega^{z}_{j}=0\) excludes row \(j\) from automated estimation, while a positive value scales its precision relative to \textsf{Full}.
A current estimate becomes authoritative only if the authorized measurement model has rank 29, positive residual degrees of freedom, an accepted information-quality gate, and a passing residual bad-data test.
If a gate fails, the previous valid estimate may be used for one declared stale epoch.
A further consecutive failure enters \textsf{Uncertified-hold} and prohibits new estimator-dependent commands.
Appendix~\ref{app:ieee30-formulation-details} specifies the matrix construction, \ac{wls}, residual covariance, bad-data test, and fallback equations.

\textbf{Dispatch and command enforcement.}
Let \(\widetilde{\mathbf{x}}_{k}\in\mathbb{R}^{29}\) be the accepted current estimate or permitted one-epoch stale estimate, \(\mathbf{p}^{0}_{k}\in\mathbb{R}^{6}\) the current validated generator-output vector, and \(P^{\mathrm{agg}}_{d,k}\in\mathbb{R}_{\geq0}\) the protected aggregate demand forecast.
The deterministic corrective-dispatch mapping \(\mathcal{K}\) infers a spatial demand pattern from \(\widetilde{\mathbf{x}}_{k}\), rescales it to \(P^{\mathrm{agg}}_{d,k}\), and solves the constrained dispatch problem using the controller-relevant decision context \(\boldsymbol{\xi}_{k}\).
It produces the single dispatch request
\begin{equation}
\mathcal{D}_{k}
=\mathcal{K}\!\left(
\widetilde{\mathbf{x}}_{k},
\mathbf{p}^{0}_{k},
P^{\mathrm{agg}}_{d,k},
\boldsymbol{\xi}_{k}
\right),
\label{eq:ieee30-dispatch-request}
\end{equation}
where \(\mathcal{D}_{k}\) is reused for candidate prediction, enforcement, counterfactual comparison, and logging at epoch \(k\).
Appendix~\ref{app:ieee30-dispatch} gives the dispatch optimization and the complete definition of \(\mathcal{D}\_{k}\).
The realized load is unavailable to the selector.
After mode-specific enforcement, balancing may adjust only \textsf{Full} and \textsf{Guarded} generators, while remaining deficits may use bounded cumulative emergency shedding.

\textbf{Admissible configurations and bounded selection.}
The belief \(\mathcal{B}_{k}=\{\mathbf{p}^{z}_{k},\mathbf{p}^{u}_{k}\}\) contains the estimated compromise probabilities of the telemetry and command interfaces.
The decision-time context \(\boldsymbol{\xi}_{k}\) contains the information available to the selector and controller, including the quantities in Eq.~\eqref{eq:ieee30-dispatch-request}, controller headroom, preceding modes, and declared uncertainty.
The implementation first selects telemetry modes through a bounded greedy search, acquires and validates the current sample, and then enumerates exactly the admissible command vectors within Hamming distance two of the preceding six-generator vector, a neighborhood of at most 154 vectors.
Each command candidate is screened before ranking for dispatch success, generation and shedding bounds, active-power balance, the synthetic hard branch ceiling, and declared hard dynamic events.
If none is feasible, the declared fallback holds the previous vector and records \textsf{No-feasible-candidate}.

\textbf{Safety-Engine terms.}
The staged implementation preserves the four quantities defined in Section~\ref{subsec:general-safety-rule}, while evaluating only those that vary at each stage.
For telemetry, residual cyber risk is
\begin{equation}
    R^{z}_{\mathrm{cyb},k}
    =\alpha_{\mathrm{spr}}R_{\mathrm{spr},k}
    +\alpha_{\mathrm{FDI}}R_{\mathrm{FDI},k}
    +\alpha_{\mathrm{raw}}R_{\mathrm{raw},k},
\label{eq:ieee30-telemetry-cyber-risk}
\end{equation}
where \(R_{\mathrm{spr},k}\) measures predicted compromise propagation under the peer authority retained by the candidate; \(R_{\mathrm{FDI},k}\) measures retained attack-induced state displacement weighted by bad-data bypass probability over the fixed attack bank; and \(R_{\mathrm{raw},k}\) measures compromise-weighted raw-ingress exposure.
The nonnegative coefficients \(\alpha_{\mathrm{spr}}\), \(\alpha_{\mathrm{FDI}}\), and \(\alpha_{\mathrm{raw}}\) weight these normalized components.
Their complete construction is given in Appendix~\ref{app:ieee30-formulation-details}.
The corresponding telemetry operational cost is
\begin{equation}
    C^{z}_{\mathrm{op},k}
    =\beta_{v}C_{\mathrm{vis},k}
    +\beta_{I}C_{\mathrm{info},k},
\label{eq:ieee30-telemetry-operational-cost}
\end{equation}
where \(C_{\mathrm{vis},k}\) is the fraction of raw telemetry made unavailable and \(C_{\mathrm{info},k}\) is the clipped degradation of the inverse-information trace relative to the all-\textsf{Full} estimator.
The nonnegative coefficients \(\beta_v\) and \(\beta_I\) weight the two normalized components.
The information term is a policy-dependent estimation-quality proxy, not a physical sensor-noise covariance or a measure of attack bias.

For command candidate \(\boldsymbol{\sigma}^{u}\), the retained cyber risk is
\begin{equation}
    R^{u}_{\mathrm{cyb},k}(\boldsymbol{\sigma}^{u})
    =\frac{1}{|\mathcal{G}|}
    \sum_{g\in\mathcal{G}}
    p^{u}_{g,k}\,a_{g}(\sigma^{u}_{g}),
\label{eq:ieee30-command-cyber-risk}
\end{equation}
where \(p^{u}_{g,k}\) is the estimated compromise probability of command interface \(g\), and \(a_g\) is its retained central-command authority: one for \textsf{Full}, the declared retained fraction for \textsf{Guarded}, and zero for \textsf{Safe-local} and \textsf{Isolated}.
The static consequence score is
\begin{equation}
    \begin{aligned}
    C^{\mathrm{stat}}_{k}(\boldsymbol{\sigma}^{u})
    &=\gamma_f C_{f,k}(\boldsymbol{\sigma}^{u})
    +\gamma_s C_{s,k}(\boldsymbol{\sigma}^{u}) \\
    &+\gamma_b C_{b,k}(\boldsymbol{\sigma}^{u})
    +\gamma_h\mathbf{1}_{\mathcal{H}^{\mathrm{stat}}_{k}(\boldsymbol{\sigma}^{u})},
    \end{aligned}
\label{eq:ieee30-physical-consequence}
\end{equation}
where \(C_{f,k}\), \(C_{s,k}\), and \(C_{b,k}\) are the normalized normal-envelope branch exceedance, total modeled shedding, and unresolved active-power imbalance, respectively.
The indicator \(\mathbf{1}_{\mathcal{H}^{\mathrm{stat}}_{k}(\boldsymbol{\sigma}^{u})}\) equals one when the event occurs and zero otherwise.
It is therefore zero for candidates admitted to ordinary ranking, but records hard violations in counterfactual and diagnostic evaluations.
The nonnegative coefficients \(\gamma_f\), \(\gamma_s\), \(\gamma_b\), and \(\gamma_h\) weight these components, and the indicator records whether the candidate violates any declared static hard condition.
Appendix~\ref{app:ieee30-formulation-details} gives the complete definitions.
Crossing a normal branch rating \(F_{\ell}\) incurs finite cost, whereas crossing the synthetic hard ceiling \(1.15F_{\ell}\) makes the candidate inadmissible.
The hard indicator is therefore zero for ranked candidates but records violations in counterfactual and diagnostic evaluations.

The static response-induced term compares each candidate with an all-\textsf{Full} counterfactual evaluated from the same initial state, estimate, forecast, dispatch request, and uncertainty node:
\begin{equation}
    R_{\mathrm{stat},k}(\boldsymbol{\sigma}^{u})
    =\left[
    C^{\mathrm{stat}}_{k}(\boldsymbol{\sigma}^{u})
    -C^{\mathrm{stat}}_{k}(\boldsymbol{\sigma}^{u,\mathrm{F}})
    \right]_{+},
\label{eq:ieee30-static-response-risk}
\end{equation}
where \(\boldsymbol{\sigma}^{u,\mathrm{F}}\) is the all-\textsf{Full} command configuration and \([x]_{+}=\max\{x,0\}\).
In a configuration with dynamic and timing-aware enforcement, we add the paired expected positive dynamic increment \(R_{\mathrm{dyn},k}\), so that
\begin{equation}
    R_{\mathrm{resp},k}
    =R_{\mathrm{stat},k}+R_{\mathrm{dyn},k}.
\label{eq:ieee30-response-risk}
\end{equation}
The selector evaluates \(R_{\mathrm{dyn},k}\) over frozen forecast and uncertainty nodes.
Sampled episode jitter and the realized load trajectory are reserved for post-selection assessment.
The dynamic term ranks continuous short-horizon recovery stress, but its 0.5-Hz frequency-deviation and 0.5-Hz/s rate of change of frequency (RoCoF) scales are reference values rather than hard limits.
Hard dynamic rejection is limited to numerical instability, frequency deviation beyond the declared synthetic 1.5-Hz bound, or excessive under-frequency shedding.
No hard RoCoF limit is claimed.

Switching costs are the normalized mode movements
\begin{equation}
    \begin{aligned}
        C^{z}_{\mathrm{sw},k}
        &=\frac{1}{3|\mathcal{E}^{z}|}
        \sum_{e\in\mathcal{E}^{z}}
        |\sigma^{z}_{e,k}-\sigma^{z}_{e,k-1}|,\\
        C^{u}_{\mathrm{sw},k}
        &=\frac{1}{3|\mathcal{G}|}
        \sum_{g\in\mathcal{G}}
        |\sigma^{u}_{g,k}-\sigma^{u}_{g,k-1}|.
    \end{aligned}
\label{eq:ieee30-switching-cost}
\end{equation}
Dwell and restoration requirements remain explicit transition rules rather than soft penalties.
The exact staged objectives, switching normalizations, validity gates, counterfactual coupling, and frozen parameters are reported in Appendix~\ref{app:ieee30-formulation-details}.
This instantiation should demonstrate that capability separation and cyber--physical enforcement trade-offs can be implemented and audited, but it does not establish certified safety, universal performance improvement, or global optimality.

%% file: Sections/7-Evaluation.tex
\section{Experimental Evaluation} \label{sec:ieee30-experimental-evaluation}
This section evaluates the IEEE 30-bus instantiation introduced in Sec.~\ref{sec:ieee30-safety-engine-instantiation}.
It examines whether capability-specific enforcement can be executed and audited, whether restriction-induced consequences can be measured, and when the implemented response terms affect selection.
We make the implementation and evaluation source code available on our GitHub repository.~\footnote{\url{https://github.com/aleLtt/SafetyAware-ZeroTrust}}

\input{Tables/Configurations}

\subsection{Setup and Implementation} \label{subsec:ieee30-implementation-setup}
\textbf{Setup.}
The evaluation uses a Python simulation of the 100-MVA \texttt{case\_ieee30} network distributed with \textit{pandapower}~3.4.0 and derived from MATPOWER data~\cite{MATPOWER,Pandapower2018}.
Corrective dispatch is formulated as a linear program enforcing active-power balance, generator bounds, per-epoch flexibility, bounded load shedding, and controller branch headroom, and is solved using SciPy's \texttt{linprog} with the HiGHS method~\cite{SciPy2020,HiGHS2018}.
Appendix~\ref{app:ieee30-dispatch} gives the complete formulation.

The implementation separates the cyber process, telemetry mediation, estimation, \ac{se} selection, command enforcement, and plant evaluation into distinct modules.
It applies the telemetry, command semantics and estimator-validity gates defined in Sec.~\ref{sec:ieee30-safety-engine-instantiation}.
Physical sensor noise remains separate from policy precision, and only a current accepted estimate or the declared one-epoch stale estimate may support a new command.
Selection and dispatch use forecast demand, while the realized load is withheld until plant evaluation.
Exactly one dispatch request is generated per epoch and reused for candidate prediction, enforcement, counterfactual comparison, and logging.

The dynamic and timing-aware command enforcement mode evaluates dynamic response prediction and mode-dependent authorization delay.
Its selector accounts for authorization jitter through a three-node Gaussian quadrature but does not observe the subsequently realized jitter.
Activation events are inserted at their realized times in the integration grid, and each decision is assessed through an independent five-second dynamic snapshot initialized from the current generation state.

\textbf{Staged implementation.}
Table~\ref{tab:ieee30-stage-summary} defines the validation configurations and the mechanism added at each evaluation stage.
\textit{Graduated Telemetry} examines the effects of capability-specific telemetry control on containment and information availability; \textit{Static Consequence-Aware} adds quasi-static command enforcement; and \textit{Dynamic and Timing-Aware} adds dynamic response prediction and mode-dependent authorization delay.
The \emph{Frozen Link} and \emph{Clean Model} configurations validate the reconstructed cyber topology and attack-free physical pipeline, respectively, while \emph{No Enforcement} and \emph{Binary Isolation} provide the \textit{Graduated Telemetry} references.
Exact seeds, campaign sizes, coefficients, and numerical parameters are reported in Appendix~\ref{app:ieee30-campaign-parameters}.

All campaigns follow the causal order defined in Sec.~\ref{subsec:ieee30-system-model}.
At each epoch, telemetry is selected before the current sample; an accepted or permitted fallback estimate then supports the single dispatch request and bounded command search.
Hard conditions exclude inadmissible candidates before objective ranking.
Realized load and execution jitter enter only post-selection enforcement and plant evaluation, whose outputs update the next operating context and defender belief.

\subsection{Evaluation Goals, Methodology, and Metrics} \label{subsec:ieee30-evaluation-methodology}
\textbf{Attack scenario.}
The evaluation combines compromise propagation with topology-consistent \ac{fdia}.
Each episode begins with a limited set of compromised telemetry endpoints.
Compromise spreads or recovers over the reconstructed peer graph, and noisy alerts update the defender-visible belief.
At each epoch, an attack direction \(\mathbf{H}\mathbf{c}_{k}\), drawn from a fixed bank and scaled relative to the measurement-noise norm, is masked by endpoint ownership so that only measurements belonging to compromised endpoints can be modified.
The resulting observation is processed by the \ac{wls} estimator and bad-data detector.

\acp{fdia} are relevant because state estimation converts field measurements into the operating context used by automated control.
Carefully structured false measurements can bias the inferred state while remaining consistent with residual-based detection~\cite{LIU2011}.
Stuxnet illustrates the broader cyber-physical integrity pattern in which malicious control influence and manipulated process feedback interact, although it is not equivalent to the state-estimation attack modeled here~\cite{LANGNER2011}.
Topology-consistent \ac{fdia} is also the attack class considered by CP-\ac{zta}~\cite{CP-ZTA}; retaining it links the reconstructed cyber benchmark to the capability-specific Safety-Engine evaluation.

\textbf{Evaluation goals.}
\textit{Graduated Telemetry} asks how unrestricted, binary, and graduated telemetry enforcement trade containment against raw visibility, structural observability, and estimation quality.
The policies receive identical compromise, alert, attack, and measurement-noise realizations, so their differences are attributable to enforcement.

\textit{Static Consequence-Aware} asks whether command restriction creates a modeled physical consequence and whether \(R_{\mathrm{stat}}\) changes selection.
The response-aware selector is paired with an ablation that removes only \(R_{\mathrm{stat}}\), retaining the same candidate set, hard screen, and other objective priorities.
The comparison is repeated at demand multipliers \(L\in\{1.00,1.05,1.10\}\).
Both selectors receive the same cyber evidence, estimate, forecast, dispatch request, and pre-decision plant state.

\textit{Dynamic and Timing-Aware} compares the \textit{Static Consequence-Aware} selector with one that adds \(R_{\mathrm{dyn}}\) at \(L=1.10\) and \(S_F=1.70\), first under default timing and then under a slower-delay configuration.
Candidate and matched all-\textsf{Full} trajectories share the same frozen quadrature nodes, while realized jitter remains unavailable until after selection.
The paired campaigns therefore test the incremental decision effect of dynamic information and the realized consequence of authorization delay without granting the selector future information.

\textbf{Metrics.}
Table~\ref{tab:ieee30-metric-meaning} defines the reported metrics.
The complete command vector is compared because identical mode fractions assigned to different generators may produce different dispatch and flow outcomes.
The trajectory-stress score combines normalized frequency deviation, RoCoF, settling duration, and terminal error.

\input{Tables/Metrics}

\subsection{Model Validation} \label{subsec:ieee30-pre-evaluation-validation}
Before comparing policies, we verify finite model constraints, attack-free feasibility, and the declared causal and enforcement invariants.
These checks establish implementation consistency, not policy superiority.

\textbf{Validation configurations.}
The \emph{No Enforcement} configuration retains all 83 peer links and verifies the unrestricted cyber and telemetry path.
The \emph{Frozen Link} reconstruction validates the structural semantics of the deterministic CP-\ac{zta} link-denial problem~\cite{CP-ZTA}.
Specifically, the graph contains 55 nodes and 83 links and each policy removes exactly its 10- or 15-link budget, denial affects only peer authority, and fixed-seed runs reproduce the same selections and metrics.
We verify that propagation and impactful bypass change in the expected direction, without comparing exact percentages, because the experiment reconstructs the topology and link-denial semantics rather than the original attack and calibration pipeline.

The \emph{Binary Isolation} configuration validates \textsf{Full}/\textsf{Isolated} endpoint enforcement and supplies the coarse \textit{Graduated Telemetry} baseline.
The attack-free \emph{Clean Model} configuration uses all-\textsf{Full} telemetry and commands to validate estimation, dispatch, plant, and dynamic paths independently of attack or policy selection.

\textbf{Benchmark calibration.}
Finite limits replace conversion-layer sentinel bounds for all six generators.
Because \texttt{case\_ieee30} lacks finite branch ratings, the \texttt{RATE\_A} pattern from the companion MATPOWER \texttt{case30} instance is mapped onto the common topology.
Controller headroom is calibrated under attack-free all-\textsf{Full} operation from positive actual-minus-predicted flow error using branch-specific 0.99 quantiles and a system-wide 0.99-quantile floor.
The controller uses the resulting reduced internal envelope, while normal-envelope and synthetic hard events remain distinct reporting categories.
Calibration covers
\begin{equation}
\mathcal{L}={1.00,1.05,1.10},
\qquad
\mathcal{S}_{F}={1.70,1.90,2.10},
\label{eq:ieee30-calibration-family}
\end{equation}
where \(L\) scales demand and \(S_F\) scales the augmented branch envelope without changing topology.
The reference point is selected for proximity to a 0.75 loading-stress target, which provides material loading without beginning from a violation.
Attack success and policy performance do not enter this selection.

\subsection{Experimental Results} \label{subsec:ieee30-results}
Tables~\ref{tab:ieee30-stage1-results},~\ref{tab:ieee30-stage2-results}, and \ref{tab:ieee30-stage3-results} report the stage-specific evidence.

\textbf{\textit{Graduated Telemetry}: containment-information trade-off.}
Relative to \textit{No Enforcement}, \textit{Graduated Telemetry} enforcement reduces mean compromised-endpoint fraction by 84.2\% and impactful \ac{fdia} bypass by 91.3\%.
\textit{Binary Isolation} provides slightly stronger containment, observability, and estimation-quality results, whereas \textit{Graduated Enforcement} retains more raw visibility (0.98568 versus 0.98110).
Thus, capability separation exposes a measurable choice between stronger endpoint removal and retaining telemetry for authorized observation.

\input{Tables/Stage1_results}

\textbf{\textit{Static Consequence-Aware}: physical consequence with a null selector effect.}
The response-aware and response-blind selectors choose the same six-generator vector in every paired epoch, and the selected static response term is never positive.
The implemented \(R_{\mathrm{stat}}\) therefore does not affect selection in the tested scenario family.
Nevertheless, the positive response-induced cost at every load shows that restricting command authority can worsen the modeled static outcome relative to applying the same request through all-\textsf{Full} interfaces.
The restricted policy produces 15, 18, and 23 infeasible epochs as load increases, while the matched all-\textsf{Full} reference produces none.
At \(L=1.10\), the events are unresolved generation-surplus violations involving at least one \textsf{Safe-local} generator.
\textit{Static Consequence-Aware} therefore demonstrates restriction-induced physical consequences, but not a prospective benefit from the static response term.

\input{Tables/Stage2_results}

\textbf{\textit{Dynamic and Timing-Aware}: sparse selector changes and mixed realized effects.}
The dynamic term is positive in 4,449 of 8,000 epochs but changes the exact command vector in only 92 (1.15\%).
Among those 92 decisions, \textit{Dynamic and Timing-Aware} produces lower realized trajectory stress in 40 cases, higher stress in 49, and equal stress in three.
Aggregate frequency-deviation and RoCoF metrics remain nearly unchanged, and the small reductions in static infeasibility and mean static cost have paired intervals that include zero.
The dynamic term is therefore decision-relevant in a small subset of cases, but its realized effects are mixed rather than consistently beneficial.

\input{Tables/Stage3_results}

\textbf{Authorization-timing sensitivity.}
Under the timing-blind \textit{Static Consequence-Aware} selector, default and slower authorization produce identical command vectors and static outcomes, isolating the effect of delayed execution.
Increasing the \textsf{Full}/\textsf{Guarded} delays from \(0.10/0.20\)~s to \(0.40/0.60\)~s raises mean maximum frequency deviation from 0.02789 to 0.04216~Hz and mean settling time from 0.07740 to 0.23023~s; the paired 95\% intervals for both increases exclude zero.
Mean maximum RoCoF remains essentially unchanged.
Thus, slower authorization allows the trajectory to move farther from nominal and delays recovery without materially changing its initial rate of change.

The timing-aware \textit{Dynamic and Timing-Aware} selector changes 79 of 8,000 command vectors under the slower delays, confirming that authorization timing can enter capability selection but only sparsely in this campaign.
No timing configuration produces a declared hard dynamic event, and the associated static and predicted-risk differences do not establish systematic improvement.
The timing experiment therefore demonstrates measurable recovery stress and limited decision sensitivity, not universal degradation bounds or a consistently superior timing-aware policy; broader implications are discussed in Sec.~\ref{sec:discussion}.

%% file: Tables/Configurations.tex
\begin{table*}[!t]
\centering
\renewcommand{\arraystretch}{1.20}
\caption{Validation and staged evaluation design.}
\label{tab:ieee30-stage-summary}
\resizebox{\textwidth}{!}{%
\begin{tabular}{l|c|c|c|c}
\hline
\rowcolor[HTML]{C0C0C0}
\multicolumn{1}{c|}{\cellcolor[HTML]{C0C0C0}\textbf{Configuration}} &
\textbf{Role} &
\textbf{Supported modes} &
\begin{tabular}[c]{@{}c@{}}\textbf{Implemented}\\ \textbf{Safety-Engine terms}\end{tabular} &
\textbf{Purpose} \\ \hline

\textit{No Enforcement} &
\begin{tabular}[c]{@{}c@{}}Validation\end{tabular} &
All telemetry endpoints unrestricted. &
\begin{tabular}[c]{@{}c@{}}N.A.\\(no \ac{se} selection)\end{tabular} &
Unrestricted cyber and telemetry reference. \\ \hline

\rowcolor[HTML]{EFEFEF}
\textit{Frozen Link} &
Validation &
Static denial of 10 or 15 peer links. &
\begin{tabular}[c]{@{}c@{}}N.A.\\(no \ac{se} selection)\end{tabular} &
Historical topology and link-denial validation. \\ \hline

\textit{Binary Isolation} &
\begin{tabular}[c]{@{}c@{}}Validation\end{tabular} &
Telemetry: \textsf{Full} or \textsf{Isolated}. &
\begin{tabular}[c]{@{}c@{}}N.A.\\(binary enforcement)\end{tabular} &
Coarse capability-removal baseline for \textit{Graduated Telemetry}. \\ \hline

\rowcolor[HTML]{EFEFEF}
\textit{Clean Model} &
Validation &
\begin{tabular}[c]{@{}c@{}}All-\textsf{Full} telemetry and commands\\ attack-free operation.\end{tabular} &
\begin{tabular}[c]{@{}c@{}}N.A.\\(fixed configuration)\end{tabular} &
Validate estimation, dispatch, plant, and dynamics. \\ \hline

\textit{Graduated Telemetry} &
Evaluation &
\begin{tabular}[c]{@{}c@{}}Telemetry: \textsf{Full}, \textsf{Guarded},\\ \textsf{Observation-only}, \textsf{Isolated}.\end{tabular} &
\(R_{\mathrm{cyb}}\), \(C_{\mathrm{op}}\), \(C_{\mathrm{sw}}\). &
Characterize containment information trade-offs. \\ \hline

\rowcolor[HTML]{EFEFEF}
\textit{Static Consequence-Aware} &
Evaluation &
\begin{tabular}[c]{@{}c@{}}Telemetry: \textit{Graduated Telemetry}; \\Commands: \textsf{Full}, \textsf{Guarded}, \textsf{Safe-local}, \textsf{Isolated}.\end{tabular} &
\begin{tabular}[c]{@{}c@{}}\(R_{\mathrm{cyb}}\), \(C_{\mathrm{op}}\), \(C_{\mathrm{sw}}\),\\\(R_{\mathrm{resp}}=R_{\mathrm{stat}}\).\end{tabular} &
Test static consequence and selector effects. \\ \hline

\textit{Dynamic and Timing-Aware} &
Evaluation &
\begin{tabular}[c]{@{}c@{}}\textit{Static Consequence-Aware} modes \\with activation delays.\end{tabular} &
\begin{tabular}[c]{@{}c@{}}\(R_{\mathrm{cyb}}\), \(C_{\mathrm{op}}\), \(C_{\mathrm{sw}}\),\\\(R_{\mathrm{resp}}=R_{\mathrm{stat}}+R_{\mathrm{dyn}}\).\end{tabular} &
Test dynamic selector and authorization-timing effects. \\ \hline
\end{tabular}%
}
\end{table*}

%% file: Tables/Metrics.tex
\begin{table*}[!t]
\centering
\caption{Evaluation metrics.}
\label{tab:ieee30-metric-meaning}

\newcommand{\rowpad}{\rule[-3pt]{0pt}{26pt}}

\resizebox{\textwidth}{!}{%
\begin{tabular}{l|l|l}
\rowcolor[HTML]{C0C0C0} 
\hline
\multicolumn{1}{c|}{\cellcolor[HTML]{C0C0C0}\textbf{\begin{tabular}[c]{@{}c@{}}Evaluation\\ Stage\end{tabular}}} &
  \multicolumn{1}{c|}{\cellcolor[HTML]{C0C0C0}\textbf{Metric}} &
  \multicolumn{1}{c}{\cellcolor[HTML]{C0C0C0}\textbf{Interpretation}} \\ \hline
 &
  Mean compromised-endpoint&
  Mean fraction of telemetry endpoints that are compromised across the evaluated epochs. \\
 &
  \begin{tabular}[c]{@{}l@{}}Impactful FDIA bypass rate\end{tabular} \rowpad&
  \begin{tabular}[c]{@{}l@{}}Fraction of FDIAs passing bad-data detection retaining at least 5\% of their state displacement.\end{tabular} \\
 &
  \begin{tabular}[c]{@{}l@{}}Raw-ingress exposure\end{tabular} \rowpad&
  \begin{tabular}[c]{@{}l@{}}Compromise-probability-weighted telemetry admitted to the trusted processing path\\(averaged across epochs).\end{tabular} \\
 &
  Raw-visibility ratio \rowpad&
  \begin{tabular}[c]{@{}l@{}}Mean fraction of telemetry streams available for authorized observation.\end{tabular} \\
\multirow{-12}{*}{\begin{tabular}[l]{@{}l@{}}\textit{Graduated}\\ \textit{Telemetry}\end{tabular}} \rowpad&
  \begin{tabular}[c]{@{}l@{}}Inverse-information \\trace score\end{tabular} &
  \begin{tabular}[c]{@{}l@{}}Mean ratio of the authorized estimator’s inverse-information trace to the all-Full reference. \\Unobservable epochs receive the declared penalty of 100.\end{tabular} \\ \hline

 & 
  \begin{tabular}[c]{@{}l@{}}Selected \(R_{stat}\) activation\end{tabular}&
  \begin{tabular}[c]{@{}l@{}}Number of epochs in which the selected command configuration has \(R_{\mathrm{stat}}>0\), over the \\total number of decisions.\end{tabular} \\
 &
  \begin{tabular}[c]{@{}l@{}}Response-aware/blind \\selection differences\end{tabular} \rowpad&
  \begin{tabular}[c]{@{}l@{}}Number of epochs in which the response-aware and response-blind selectors choose different \\complete command vectors.\end{tabular} \\
 
 \multirow{-4}{*}{\begin{tabular}[l]{@{}l@{}}\textit{Static}\\ \textit{Consequence-Aware}\end{tabular}}&
  \begin{tabular}[c]{@{}l@{}}Mean response-induced cost\end{tabular} \rowpad&
  \begin{tabular}[c]{@{}l@{}}Mean positive static-cost increment of the selected configuration over the matched all-\textsf{Full} \\application of the same dispatch request.\end{tabular} \\
  
   &
  \begin{tabular}[c]{@{}l@{}}Realized static infeasibility\end{tabular} \rowpad&
  \begin{tabular}[c]{@{}l@{}}Number and rate of post-selection epochs that violate at least one declared static hard condition.\end{tabular} \\ \hline


 & 
  \begin{tabular}[c]{@{}l@{}}Predicted-\(R_{\mathrm{dyn}}\) activation\end{tabular}&
  \begin{tabular}[c]{@{}l@{}}Number of decisions for which the selected configuration has \(R_{\mathrm{dyn}}>0\).\end{tabular} \\
  &
 Selection differences \rowpad&
  \begin{tabular}[c]{@{}l@{}}Number of epochs in which \textit{Static Consequence-Aware} and \textit{Dynamic and Timing-Aware} select\\ different complete command vectors.\end{tabular} \\
  

 &
  \begin{tabular}[c]{@{}l@{}}Stress comparison in \\differing selections\end{tabular} \rowpad&
  \begin{tabular}[c]{@{}l@{}}Numbers of differing-selection epochs in which \textit{Dynamic and Timing-Aware} produces lower,\\ higher, or equal realized trajectory stress relative to \textit{Static Consequence-Aware}.\end{tabular} \\

  \multirow{-3}{*}{\begin{tabular}[l]{@{}l@{}}\textit{Dynamic and}\\ \textit{Timing-Aware}\end{tabular}} &
  Mean peak frequency deviation \rowpad&
  \begin{tabular}[c]{@{}l@{}}Mean across dynamic snapshots of the maximum absolute frequency deviation.\end{tabular} \\

  &
  Mean peak RoCoF \rowpad&
  \begin{tabular}[c]{@{}l@{}}Mean across dynamic snapshots of the maximum absolute rate of change of frequency.\end{tabular} \\



  &
  Declared hard dynamic events \rowpad&
  \begin{tabular}[c]{@{}l@{}}Number and rate of snapshots that violate at least one declared hard dynamic condition.\end{tabular} \\

\end{tabular}%
}
\end{table*}

%% file: Tables/Stage1_results.tex
\begin{table}[!t]
\centering
\renewcommand{\arraystretch}{1.3}
\caption{\textit{Graduated Telemetry} containment and information results.}
\label{tab:ieee30-stage1-results}
\resizebox{\columnwidth}{!}{%
\begin{tabular}{lccc}
\hline
\rowcolor[HTML]{C0C0C0}
\multicolumn{1}{c}{\cellcolor[HTML]{C0C0C0}\textbf{Metric}} &
\multicolumn{1}{c}{\cellcolor[HTML]{C0C0C0}\textit{\textbf{No Enforcement}}} &
\multicolumn{1}{c}{\cellcolor[HTML]{C0C0C0}\textit{\textbf{Binary Isolation}}} &
\textit{\textbf{\textit{Graduated Telemetry}}} \\ \hline
Mean compromised-endpoint              & 0.0899816     & 0.0131657 & 0.0142293 \\
Impactful \ac{fdia} bypass rate        & 0.6992654     & 0.0499929 & 0.0609656 \\
Raw-ingress exposure                   & unrestricted & 0.0033677 & 0.0064184 \\
Raw-visibility ratio                   & 1.0000000     & 0.9810966 & 0.9856750 \\
Inverse-information trace score        & 1.0000000     & 4.0956594 & 4.3267149 \\   \hline
\end{tabular}%
}
\par
\vspace{2mm}
\noindent
\parbox{\columnwidth}{\footnotesize\emph{Note:} The \textsf{Frozen Link} configuration is excluded because peer-link denial changes propagation connectivity, whereas endpoint enforcement may also change raw visibility and estimator influence; combining them would conflate distinct capabilities.}
\end{table}

%% file: Tables/Stage2_results.tex
\begin{table}[!t]
\centering
\renewcommand{\arraystretch}{1.25}
\caption{\textit{Static Consequence-Aware} consequence activation and static selector effect. Each load contains 8,000 paired decisions.}
\label{tab:ieee30-stage2-results}
\resizebox{.5\textwidth}{!}{%
\begin{tabular}{c|c|c|c|c}
\hline
\rowcolor[HTML]{C0C0C0}
\textbf{Load} &
\begin{tabular}[c]{@{}c@{}}\textbf{Selected} \(R_{\mathrm{stat}}\)\\ \textbf{activation}\end{tabular}&
\begin{tabular}[c]{@{}c@{}}\textbf{Response-aware/blind}\\ \textbf{selection differences}\end{tabular} &
\begin{tabular}[c]{@{}c@{}}\textbf{Mean response}\\ \textbf{induced cost}\end{tabular} &
\begin{tabular}[c]{@{}c@{}}\textbf{Realized static}\\ \textbf{infeasibility}\end{tabular} \\ \hline

1.00 & 0/8000 & 0/8000 & 0.00197277 & 15/8000 (0.1875\%) \\
1.05 & 0/8000 & 0/8000 & 0.00240047 & 18/8000 (0.2250\%) \\
1.10 & 0/8000 & 0/8000 & 0.00308434 & 23/8000 (0.2875\%) \\ \hline

\end{tabular}%
}
\end{table}

%% file: Tables/Stage3_results.tex
\begin{table*}[!t]
\centering
\renewcommand{\arraystretch}{1.25}
\caption{\textit{Dynamic and Timing-Aware} incremental dynamic-term results under default timing. Dynamic calculations are independent five-second decision snapshots.}
\label{tab:ieee30-stage3-results}
\resizebox{0.92\textwidth}{!}{%
\begin{tabular}{l|c|c|c}
\hline
\rowcolor[HTML]{C0C0C0}
\textbf{Metric} &
\textbf{\textit{Static Consequence-Aware}} &
\textbf{\textit{Dynamic and Timing-Aware}} &
\textbf{Observed result} \\ \hline
Predicted-\(R_{\mathrm{dyn}}\) activation & N.A. & 4449/8000 & Dynamic term is frequently active. \\

Selection differences & \multicolumn{2}{c|}{92/8000} & Sparse incremental selector effect. \\

Stress comparison in differing selections & --- & 40 lower / 49 higher / 3 equal & Realized effects are mixed. \\

Mean peak frequency deviation & 0.0278859~Hz & 0.0278865~Hz & Nearly unchanged in aggregate. \\

Mean peak RoCoF & 0.1558111~Hz/s & 0.1558185~Hz/s & Nearly unchanged in aggregate. \\



Declared hard dynamic events & 0/8000 & 0/8000 & No event under the synthetic hard definition. \\ \hline
\end{tabular}%
}
\end{table*}

%% file: Sections/8-Discussion.tex
\section{Discussion} \label{sec:discussion}

\subsection{Discussion of Experimental Results} \label{subsec:discussion-results}
The evaluation tests whether capability-specific \ac{zt} enforcement can be represented, executed, and audited when authorization affects both observation and control.
Across the three stages, it shows that peer communication, raw visibility, estimator influence, command authority, and authorization timing can be treated as distinct policy dimensions rather than as consequences of a single device-wide trust state.

\textbf{Telemetry-capability evaluation.}
\textit{Graduated Telemetry} demonstrates that graduated telemetry enforcement can reduce compromise propagation and impactful \ac{fdia} bypass while retaining more raw visibility than \textit{Binary Isolation}.
However, \textit{Binary Isolation} achieves slightly stronger containment and estimation-quality results.
The contribution is therefore the explicit containment-information trade-off, not uniformly superior performance.
In particular, preserving a stream for authorized observation neither makes it trustworthy nor authorizes its use by the estimator.
The design implication is that visibility and automated influence should be enforced separately, while observability and estimation quality remain explicit validity checks.

\textbf{Command-enforcement consequence evaluation.}
\textit{Static Consequence-Aware} demonstrates that withdrawing remote command authority can itself alter the modeled physical outcome by removing coordinated redispatch capability.
The \textsf{Safe-local} mode should consequently be understood as a declared local hold or fallback, whose suitability depends on the operating state, requested dispatch, retained balancing authority, and assigned generator; it is not intrinsically safer because it is more restrictive.
The Safety Engine must therefore evaluate the complete authority vector rather than infer physical validity from a mode label.

\textit{Static Consequence-Aware} also distinguishes consequence activation from decision relevance.
The paired counterfactual analysis identifies restriction-induced costs, yet the prospective static term is never positive for the selected candidates and its removal does not change any selected command vector.
A response may therefore create a measurable consequence even when the implemented predictor does not discriminate among the admissible choices.
This motivates three separate validation questions for any response term: whether the response creates a consequence, whether the prospective model distinguishes candidates before enforcement, and whether the resulting selection changes realized outcomes.

\textbf{Dynamic response and authorization timing.}
\textit{Dynamic and Timing-Aware} shows that causal dynamic information can affect selection, but only for a small subset of the tested decisions, and the realized effects of those changes are mixed.
It therefore establishes incremental decision relevance rather than consistent physical improvement.
This distinction is important: frequent activation of a risk term does not imply frequent policy changes, and policy changes do not imply uniformly better realized trajectories.

The timing campaign shows that command permission and timely command realization are separate properties.
Applying the same selected commands later increases frequency excursion and settling time while leaving maximum RoCoF nearly unchanged.
This behavior is consistent with the modeled mechanism: the initial generation--demand mismatch determines the peak rate of change, whereas a longer activation delay extends that mismatch, allowing frequency to move farther from nominal and prolonging recovery.
Timing information changes only a small fraction of command vectors, but its physical effect is observable even when selection is unchanged.
Under slower timing, the most common selector change is from \textsf{Guarded} to \textsf{Full}, indicating that a delayed partial correction can have greater predicted consequences than applying the requested correction fully.
Changes also occur in the opposite direction and toward \textsf{Safe-local}, confirming that authorization delay does not create a fixed physical ordering among the modes.
The relevant decision object remains the complete authority vector evaluated in its current operating context.

Finally, a candidate-versus-reference term may not expose a timing degradation shared by both trajectories.
A deployment with an absolute recovery or latency requirement therefore needs a separate deadline, absolute-stress constraint, or operating-envelope gate rather than relying exclusively on incremental response risk.
Cyber evidence, retained authority, activation timing, incremental consequence, and absolute physical stress should remain distinct but interacting decision quantities.

\subsection{Answers to the Research Questions} \label{subsec:discussion-rqs}
\textbf{Answer to RQ1.}
The analysis identifies nine convergence strains.
S1-S6 amplify enterprise-IT challenges by making continuous assessment and granular enforcement harder across constrained, autonomous, heterogeneous, dynamic, and long-lived assets.
S7-S9 acquire distinctive \ac{cps} semantics because authorization delay can invalidate a control action, authentic evidence can remain physically inaccurate or operationally unsuitable, and enforcement can remove observation, control, or recovery capabilities needed by the plant.
Latency and evidence uncertainty also occur in enterprise IT, but their direct effect on plant evolution and recovery gives them different operational consequences in a \ac{cps}.
Additional resources and conventional policy scaling may mitigate S1-S6, whereas S7-S9 require authorization to represent physical time, plant state, evidence validity, and the consequences of both retaining and withdrawing capabilities.

\textbf{Answer to RQ2.}
The corresponding requirements form three dependent layers.
R1--R6 constitute an \emph{integration-feasibility layer}, determining whether \ac{zt} controls can be implemented across constrained, autonomous, heterogeneous, dynamic, and non-modifiable environments.
R7 and R8 form an \emph{operational-validity layer}, requiring timely decisions based on physically meaningful and uncertainty-aware evidence.
R9 forms a \emph{consequence-aware enforcement layer}, requiring graduated restriction and recovery that consider residual cyber risk and restriction-induced physical consequences as distinct quantities.
The layers are cumulative: consequence-aware selection cannot compensate for an unenforceable control, a missed operational deadline, or invalid evidence.
R9 therefore depends particularly on the operation-level semantics of R5, the disruption-tolerant authorization of R7, and the qualified cyber-physical evidence of R8.
Operational validity also requires distinctions that a generic availability or trust label would hide: data presence is not observability, observability is not estimation precision, reference-scale stress is not a hard event, and reduced authority is not automatically improved physical safety.

\textbf{Answer to RQ3.}
\ac{sa-zt} constrains suspicious components by representing state-changing authority, raw telemetry visibility, automated influence, and peer communication separately.
Command-side \acp{pep} restrict writes and actuation, while the \ac{tb} controls observation and automated use.
The \ac{se} excludes unsupported, undeployable, or physically inadmissible configurations before ranking the remainder according to residual cyber risk, operational degradation, response-induced consequence, and reconfiguration cost.
This structure allows command authority or automated influence to be withdrawn while selected telemetry remains available for authorized monitoring, diagnosis, or recovery.
Restoring broader authority remains a new authorization decision subject to renewed evidence and the declared recovery conditions.
The IEEE 30-bus case demonstrates that this decision structure is representable, implementable, and auditable within the declared model; it does not establish universal policy improvement.

\subsection{Implications for Cyber-Physical \ac{zt}}
The analysis and case study yield five design principles, grouped into three concerns.

\textbf{Capability semantics.}
First, policies should name observation, automated influence, actuation, and peer communication separately, with each \ac{pep} enforcing only its assigned capability.
Second, residual adversarial risk, lost operational capability, restriction-induced consequence, and absolute plant stress should be computed and logged separately.
Although these quantities interact in selection, collapsing them into a single trust score would obscure whether a decision was driven by suspected compromise, insufficient information, reduced control authority, or physical response.

\textbf{Auditable decision structure.}
Third, hard admissibility should precede preference ranking and produce explicit reason codes.
Continuous margins may rank admissible candidates, but they should not be reported as hard failures, nor should a hard violation become acceptable because the candidate reduces cyber authority.
Fourth, each response term should be validated at the three levels exposed by the experiments: consequence activation, prospective candidate discrimination, and realized selection effect.
\textit{Static Consequence-Aware} shows that a consequence may exist without a selector effect, while \textit{Dynamic and Timing-Aware} shows that a selector effect may exist without consistent realized improvement.

\textbf{Timing, fallback, and recovery.}
Fifth, every enforcement mode should declare its activation delay, fallback behavior, validity duration, dwell conditions, restoration path, and re-authorization requirements.
The timing results show that eventual permission is insufficient: identical authority vectors can produce different trajectories depending on when they become active.
Together, these principles define the practical contribution of the \ac{se} as an auditable structure for representing cyber-physical enforcement consequences, rather than a universal safety model or power-grid controller.

\subsection{Limitations and Future Work} \label{subsec:limitations-future-work}
\textbf{Benchmark and model coverage.}
The IEEE 30-bus system is a well-established benchmark that supports reproducible power-system and cyber-physical evaluation~\cite{MATPOWER,CP-ZTA}.
Its controlled topology and operating model are suitable for demonstrating the \ac{sa-zt} decision contract, but one transmission benchmark, fixed scenario banks, and a declared attack family cannot support generalized claims across power systems or other \acp{cps}.
The DC and aggregate dynamic models capture the interactions needed for the present study, not the full range of AC, voltage, protection, cascading, communication, and adaptive-adversary effects.
Future validation should therefore vary grid topologies, operating regimes, attacks, uncertainty, and physical fidelity, including AC, protection, communication, and hardware-in-the-loop settings.
More importantly, the same contract should be instantiated in domains such as water networks, manufacturing, transportation, and building automation, where observation, actuation, degraded operation, and harm have different meanings.
This cross-domain evidence would identify which abstractions generalize and which require domain-specific reformulation.

\textbf{General formulation and guarantees.}
The present framework supplies a mathematical decision contract and an executable instantiation; the next theoretical step is to characterize the conditions under which that contract remains valid across domains.
This includes principled selection of the response baseline and assessment horizon, treatment of uncertainty and tail consequences, calibration or avoidance of scalar policy weights, and prevention of double counting between attack-induced and response-induced effects.
Future work should also identify conditions for the existence of admissible responses and derive domain-appropriate robustness, stability, invariance, or constraint-satisfaction guarantees for repeated decisions.
The objective is not to replace the framework with one universal physical model, but to specify reusable validity obligations that each domain-specific Safety Engine must satisfy.
These obligations should preserve causal information use, the separation of hard admissibility from soft ranking, explicit model-validity conditions, and declared behavior when no preferred response can be validated or installed.

\textbf{Instantiation and deployment methodology.}
The case study uses causal decomposition: bounded greedy telemetry selection followed by exact command enumeration within a Hamming-distance-two neighborhood.
This procedure does not claim joint or global optimality.
On tractable systems, comparisons with full enumeration, mixed-integer formulations, or robust and sequential alternatives can quantify the effects of decomposition and locality.
These comparisons should support a broader instantiation methodology rather than optimize only the present grid model.
Such a methodology should require each domain to declare its enforcement units and modes, evidence and timing model, hard admissibility conditions, consequence and counterfactual models, fallback and recovery procedures, and audit record.
The common framework would define the interfaces and invariants among these elements, while domain modules would supply the physical models and guarantees appropriate to the protected process.
Deployment validation must additionally cover measured activation delays, policy-version consistency, protected fallback paths, hysteresis and dwell rules, re-attestation, restoration, and operator intervention.
Reusable conformance criteria could then verify that an instantiation respects causal timing, never trades a hard violation against a favorable soft score, and reevaluates validity when runtime enforcement changes the authorized capabilities.
Establishing these criteria is the next step toward applying \ac{sa-zt} consistently across heterogeneous industrial and cyber-physical settings.

%% file: Sections/Conclusion.tex
\section{Conclusion}    \label{sec:conclusion}
In this paper, we mapped the \ac{nist} \ac{zta} tenets to nine strains distinguished \ac{iot}-amplified challenges from \ac{cps}-specific strains and yielded requirements for deployable, operationally valid, and consequence-aware controls.
We introduced \textit{Safety-Aware Zero Trust}, which extends \ac{nist} \ac{zta} with a \textit{Safety Engine} and \textit{Telemetry Broker} to separate raw telemetry visibility, automated influence, and state-changing authority while selecting admissible graduated responses.
The IEEE 30-bus instantiation demonstrated this decision contract through causal information handling, bounded selection, admissibility checks, fallback behavior, and explicit authorization timing.
The evaluation exposed a containment-information trade-off and confirmed that command restriction can produce physical consequences.
Experimental evaluation provides promising evidence that capability separation preserves useful information, response awareness can alter policy selection, and authorization timing affects operational margin.
Future work should thus extend these principles and validate them using higher-fidelity models and other cyber-physical domains.


%% file: Sections/Appendix_revised.tex

\section{Detailed IEEE 30-Bus Formulation and Parameterization}
\label{app:ieee30-formulation-details}

This appendix provides the derivation and accounting details omitted from
Section~\ref{subsec:ieee30-compact-formulation}.  It describes the staged
algorithm implemented in the case study rather than an idealized joint
optimizer over all telemetry and command configurations.  The notation also
separates physical sensor noise from policy precision, a provisional numerical
estimate from an accepted estimate, a controller request from an applied
command, and operating stress from hard infeasibility.

\subsection{DC Plant and Measurement-Matrix Derivation}

Choose an arbitrary but fixed orientation for each branch. For branch
\(\ell=(i,j)\), the corresponding row of
\(\mathbf{C}\in\{-1,0,1\}^{41\times30}\) contains \(+1\) at bus \(i\),
\(-1\) at bus \(j\), and zero elsewhere. Let \(x_{\ell}^{\mathrm{pu}}>0\)
and \(\tau_{\ell}>0\) denote the branch reactance and normalized transformer
tap ratio. The MATPOWER convention that an encoded zero tap denotes unity is
applied before constructing
\begin{equation}
\mathbf{B}_{\ell}
=\operatorname{diag}\!\left(\frac{1}{x_{\ell}^{\mathrm{pu}}\tau_{\ell}}\right).
\label{eq:app-ieee30-branch-coefficients}
\end{equation}
The implementation asserts that all transformer phase shifts are zero. A
nonzero phase shift would require affine flow and measurement offsets and is
therefore rejected rather than silently ignored.

Let \(\mathbf{T}_{r}\in\{0,1\}^{30\times29}\) be the reference-bus insertion
matrix: it maps the reduced state \(\mathbf{x}_{k}\) to the complete angle
vector \(\boldsymbol{\theta}_{k}=\mathbf{T}_{r}\mathbf{x}_{k}\), with the
bus-1 angle fixed to zero. With system base \(S_B=100\)~MVA, the MW-valued
branch flows and nodal injections satisfy
\begin{equation}
\begin{aligned}
\mathbf{f}_{k}
    &=S_B\mathbf{B}_{\ell}\mathbf{C}\mathbf{T}_{r}\mathbf{x}_{k},\\
\mathbf{C}^{\mathsf T}\mathbf{f}_{k}
    &=\mathbf{G}\mathbf{p}^{\mathrm{app}}_{k}
      -\bigl(\mathbf{d}_{k}-\mathbf{s}_{k}\bigr),
\end{aligned}
\label{eq:app-ieee30-dc-plant}
\end{equation}
where \(\mathbf{G}\in\{0,1\}^{30\times6}\) maps generator outputs to their
buses, \(\mathbf{d}_{k}\) is gross demand, and \(\mathbf{s}_{k}\) is
curtailed demand. A physical solution requires
\begin{equation}
\mathbf{1}^{\mathsf T}
\bigl(\mathbf{G}\mathbf{p}^{\mathrm{app}}_{k}
-\mathbf{d}_{k}+\mathbf{s}_{k}\bigr)=0.
\label{eq:app-ieee30-balance}
\end{equation}
The reduced DC solve is performed after recording this residual. When the
residual is nonzero, the implementation introduces a virtual slack injection
only to obtain a diagnostic angle and flow vector; the virtual term is logged
as a balance violation and does not modify generation or represent authority
at bus~1.

Three branch envelopes have distinct roles. Let \(F_{\ell}\) be the normal
operating-envelope rating, \(m_{\ell}\) the attack-independent dispatch
headroom, and
\begin{equation}
F^{\mathrm{ctrl}}_{\ell}=F_{\ell}-m_{\ell},
\qquad
F^{\mathrm{hard}}_{\ell}=1.15F_{\ell}.
\label{eq:app-ieee30-three-flow-envelopes}
\end{equation}
The controller plans inside \(F^{\mathrm{ctrl}}_{\ell}\); crossing
\(F_{\ell}\) incurs finite operating stress; and crossing the synthetic
benchmark ceiling \(F^{\mathrm{hard}}_{\ell}\) is infeasible. The factor
1.15 is not presented as a utility protection rating.

The branch-flow measurement block is
\begin{equation}
\mathbf{H}^{f}
=\mathbf{B}_{\ell}\mathbf{C}\mathbf{T}_{r}
\in\mathbb{R}^{41\times29}.
\label{eq:app-ieee30-flow-measurement-block}
\end{equation}
Let \(\mathbf{M}_{p}\in\{0,1\}^{14\times30}\) select the buses in
\(\mathcal{N}^{p}\) at which injection measurements are available. Because
per-unit nodal injections are
\(\mathbf{C}^{\mathsf T}\mathbf{H}^{f}\mathbf{x}_{k}\), the injection
measurement block is
\begin{equation}
\mathbf{H}^{p}
=\mathbf{M}_{p}\mathbf{C}^{\mathsf T}\mathbf{H}^{f}
=\mathbf{M}_{p}\mathbf{C}^{\mathsf T}
 \mathbf{B}_{\ell}\mathbf{C}\mathbf{T}_{r}
\in\mathbb{R}^{14\times29}.
\label{eq:app-ieee30-injection-measurement-block}
\end{equation}
The implementation stores measurements in logical endpoint order. If
\(\boldsymbol{\Pi}_{z}\in\{0,1\}^{55\times55}\) is the corresponding fixed
permutation matrix, then
\begin{equation}
\mathbf{H}=\boldsymbol{\Pi}_{z}
\begin{bmatrix}\mathbf{H}^{f}\\\mathbf{H}^{p}\end{bmatrix},
\qquad
\operatorname{rank}(\mathbf{H})=29.
\label{eq:app-ieee30-measurement-matrix}
\end{equation}
The construction checks the base MVA, bus~1 reference, branch endpoints and
ordering, tap ratios, zero phase shifts, full and reduced bus matrices, and the
alignment of all flow and injection rows. These checks ensure that estimation
and closed-loop propagation use the same DC model.

\subsection{Attack Realizability and Compromise Belief}

Let \(\mathbf{q}^{z}_{k}\in\{0,1\}^{55}\) denote the hidden compromise state
of the telemetry endpoints and let \(o(j)\) identify the owner of measurement
row \(j\). A state-consistent attack direction has the form
\(\mathbf{H}\mathbf{c}_{k}\). In the simulation, the attacked sample before
estimator admission is
\begin{equation}
\mathbf{z}^{\mathrm{raw}}_{k}
=\mathbf{H}\mathbf{x}_{k}+\boldsymbol{\eta}_{k}
+\operatorname{diag}\!\bigl(q^{z}_{o(1),k},\ldots,
q^{z}_{o(55),k}\bigr)\mathbf{H}\mathbf{c}_{k},
\,
\boldsymbol{\eta}_{k}\sim\mathcal{N}(\mathbf{0},\mathbf{R}_{k}).
\label{eq:app-ieee30-realizable-fdi}
\end{equation}
Telemetry modes are selected before this sample is acquired. They determine
peer propagation, raw delivery, and estimator admission/precision; they do not
retroactively change the sensor-noise draw or relabel it as policy uncertainty.
In particular, an attacked row excluded from \ac{wls} may remain visible to an
authorized observation consumer under \textsf{Observation-only}.

The hidden compromise process uses the policy-mediated peer-authority matrix
\(\mathbf{A}(\boldsymbol{\sigma}^{z}_{k})\). Its belief-side one-step
susceptible--infected--susceptible (SIS)
prediction is
\begin{equation}
\widetilde p^{z}_{j,k+1}
=p^{z}_{j,k}(1-\gamma)
+\bigl(1-p^{z}_{j,k}\bigr)
\left[1-\prod_i
\left(1-\beta A_{ij}(\boldsymbol{\sigma}^{z}_{k})p^{z}_{i,k}\right)
\right].
\label{eq:app-ieee30-sis-belief-prediction}
\end{equation}
The \ac{se} does not observe \(\mathbf{q}^{z}_{k}\). It receives
\begin{equation}
p^{z}_{j,k}
=\Pr\!\left(q^{z}_{j,k}=1\mid\mathcal{I}^{\mathrm{sec}}_{k}\right),
\label{eq:app-ieee30-telemetry-belief}
\end{equation}
where \(\mathcal{I}^{\mathrm{sec}}_{k}\) contains defender-visible security
evidence. Command-interface probabilities \(\mathbf{p}^{u}_{k}\) are derived
from the same evidence and the declared ownership/dependency map. Hidden
simulation truth is used for outcome generation, not passed to the selector.

\subsection{Policy-Mediated State Estimation}

For telemetry configuration \(\boldsymbol{\sigma}^{z}\), raw visibility and
relative estimator precision are encoded by
\(\mathbf{v}^{z}(\boldsymbol{\sigma}^{z})\in\{0,1\}^{55}\) and
\(\boldsymbol{\omega}^{z}(\boldsymbol{\sigma}^{z})\in[0,1]^{55}\),
respectively.  The implementation enforces
\(\omega^{z}_{j}>0\Rightarrow v^{z}_{j}=1\).  Define the estimator-admitted
row set
\begin{equation}
\mathcal{J}^{z}_{k}(\boldsymbol{\sigma}^{z})
=\{j:\omega^{z}_{j}(\boldsymbol{\sigma}^{z})>0\}.
\label{eq:app-ieee30-admitted-set}
\end{equation}
Let \(\mathbf{R}_{J,k}=\operatorname{diag}(\sigma^{2}_{j,k})_{j\in J}\)
be the physical sensor-noise covariance on these rows.  Policy precision enters
only the \ac{wls} matrix
\begin{equation}
\mathbf{W}_{J,k}(\boldsymbol{\sigma}^{z})
=\operatorname{diag}\!\left(
\frac{\omega^{z}_{j}(\boldsymbol{\sigma}^{z})}{\sigma^{2}_{j,k}}
\right)_{j\in J}.
\label{eq:app-ieee30-policy-weights}
\end{equation}
Thus, \textsf{Guarded} reduces relative precision; it does not decrease
physical sensor noise or guarantee a fixed bound on attack-induced state
displacement.

Using a numerically stable singular-value decomposition of
\(\mathbf{W}_{J,k}^{1/2}\mathbf{H}_{J}\), the implementation constructs
\begin{equation}
\begin{aligned}
\mathbf{P}^{\mathrm{info}}_{J,k}
&=\left(\mathbf{H}_{J}^{\mathsf T}\mathbf{W}_{J,k}
\mathbf{H}_{J}\right)^{\dagger},\\
\mathbf{A}_{J,k}
&=\mathbf{P}^{\mathrm{info}}_{J,k}
\mathbf{H}_{J}^{\mathsf T}\mathbf{W}_{J,k},\\
\widehat{\mathbf{x}}^{\mathrm{num}}_{k}
&=\mathbf{A}_{J,k}\mathbf{z}^{\mathrm{raw}}_{J,k}.
\end{aligned}
\label{eq:app-ieee30-policy-wls}
\end{equation}
The pseudoinverse permits finite diagnostics under rank loss, but such output is
not accepted for normal control.  When the estimator is full column rank,
\(\mathbf{P}^{\mathrm{info}}_{J,k}\) is the inverse weighted information used
by the policy's quality gate.  It is not, in general, the actual estimation
error covariance.  Under fixed admitted rows and zero-mean physical noise, that
covariance is
\begin{equation}
\mathbf{V}_{J,k}
=\mathbf{A}_{J,k}\mathbf{R}_{J,k}\mathbf{A}_{J,k}^{\mathsf T},
\label{eq:app-ieee30-actual-estimator-covariance}
\end{equation}
and any attack bias \(\mathbf{A}_{J,k}\mathbf{a}_{J,k}\) must be treated
separately.  Accordingly, the manuscript refers to
\(\operatorname{tr}(\mathbf{P}^{\mathrm{info}})\) as an information or
estimation-quality proxy, not as complete physical uncertainty.

The bad-data detector is calibrated from physical noise after the
policy-dependent estimator has been constructed.  With row selector
\(\mathbf{E}_{J}\), define
\begin{equation}
\mathbf{Q}_{J,k}
=\mathbf{E}_{J}-\mathbf{H}_{J}\mathbf{A}_{J,k},
\qquad
\mathbf{S}_{r,k}
=\mathbf{Q}_{J,k}\mathbf{R}_{k}\mathbf{Q}_{J,k}^{\mathsf T}.
\label{eq:app-ieee30-residual-covariance}
\end{equation}
For residual \(\mathbf{r}_{k}=\mathbf{Q}_{J,k}\mathbf{z}^{\mathrm{raw}}_{k}\),
the statistic and threshold are
\begin{equation}
J_{k}=\mathbf{r}_{k}^{\mathsf T}\mathbf{S}_{r,k}^{\dagger}\mathbf{r}_{k},
\qquad
\nu_{k}=\operatorname{rank}(\mathbf{S}_{r,k}),
\qquad
\tau_{k}=\chi^{2}_{1-\alpha,\nu_{k}}.
\label{eq:app-ieee30-bdd}
\end{equation}
Normal acceptance requires \(J_k\leq\tau_k\) and \(\nu_k\geq1\).  The FDI
selection proxy uses the associated noncentral chi-square detection
probability for the attack-induced mean residual; it does not reuse policy
precision as physical noise.

The numerical estimate becomes authoritative only if all four gates pass:
\begin{equation}
\begin{aligned}
&\operatorname{rank}(\mathbf{H}_{J})=29,
\qquad \nu_k\geq1,\\
&\frac{\operatorname{tr}(\mathbf{P}^{\mathrm{info}}_{J,k})}
   {\operatorname{tr}(\mathbf{P}^{\mathrm{info}}_{\textsf{Full},k})}
\leq\rho_{\max},
\qquad J_k\leq\tau_k.
\end{aligned}
\label{eq:app-ieee30-estimate-validity}
\end{equation}
On success, \(\widehat{\mathbf{x}}^{\mathrm{num}}_{k}\) becomes the new last
valid estimate and the regime is \textsf{Normal-current}.  On the first
consecutive failure, the command stage may use the preceding valid estimate in
\textsf{Degraded-last-valid}.  Further consecutive failure enters
\textsf{Uncertified-hold}: the controller issues a non-estimator-dependent hold
request and no new estimator-dependent command-mode search is performed.

\subsection{Command-Mode Realization}

Let \(\mathbf{p}^{0}_{k}\in\mathbb{R}^{6}\) be the current validated
generator-output vector. The single corrective-dispatch request introduced in
Eq.~\eqref{eq:ieee30-dispatch-request} is represented in the appendix as
\begin{equation}
\mathcal{D}_{k}
=\bigl(\mathbf{p}^{\mathrm{req}}_{k},
       \mathbf{s}^{\mathrm{req}}_{k},
       \widehat{\mathbf{f}}_{k},
       \mathrm{status}_{k}\bigr),
\label{eq:app-ieee30-dispatch-request-components}
\end{equation}
where \(\mathbf{p}^{\mathrm{req}}_{k}\in\mathbb{R}^{6}\) contains the
requested generator setpoints, \(\mathbf{s}^{\mathrm{req}}_{k}\in
\mathbb{R}_{\geq0}^{30}\) the scheduled bus-level shedding,
\(\widehat{\mathbf{f}}_{k}\in\mathbb{R}^{41}\) the branch flows predicted
for that request, and \(\mathrm{status}_{k}\) the corrective-dispatch solver
status. Appendix~\ref{app:ieee30-dispatch} defines their construction.

For generator \(g\), define the ordinary per-epoch envelope
\begin{equation}
\Gamma^{\mathrm{F}}_{g,k}
=\left[
\max\{p^{\min}_{g},p^{0}_{g,k}-\Delta p^{\max}_{g}\},
\min\{p^{\max}_{g},p^{0}_{g,k}+\Delta p^{\max}_{g}\}
\right]
\label{eq:app-ieee30-full-envelope}
\end{equation}
and the guarded envelope \(\Gamma^{\mathrm{G}}_{g,k}\) obtained by replacing
\(\Delta p^{\max}_{g}\) with \(0.35\Delta p^{\max}_{g}\). Before central
balance correction, the command \ac{pep} realizes
\begin{equation}
p^{\mathrm{pre}}_{g,k}
=
\begin{cases}
\Pi_{\Gamma^{\mathrm{F}}_{g,k}}(p^{\mathrm{req}}_{g,k}),
  &\sigma^{u}_{g,k}=\textsf{Full},\\
\Pi_{\Gamma^{\mathrm{G}}_{g,k}}(p^{\mathrm{req}}_{g,k}),
  &\sigma^{u}_{g,k}=\textsf{Guarded},\\
p^{0}_{g,k},
  &\sigma^{u}_{g,k}=\textsf{Safe-local},\\
p^{0}_{g,k},
  &\sigma^{u}_{g,k}=\textsf{Isolated}.
\end{cases}
\label{eq:app-ieee30-command-realization}
\end{equation}
The identical quasi-static setpoint in the last two cases is deliberate.
\textsf{Safe-local} denotes a local hold/protection fallback, whereas
\textsf{Isolated} additionally removes the ordinary remote interface. Their
network, transition, and timing semantics differ; a distinct hidden balancing
law is not assigned to \textsf{Safe-local}.

Let \(\mathcal{G}^{\mathrm{bal}}_{k}\) contain only generators in
\textsf{Full} or \textsf{Guarded}. Central balance correction must satisfy
\begin{equation}
\Delta p^{\mathrm{bal}}_{g,k}=0
\quad\forall g\notin\mathcal{G}^{\mathrm{bal}}_{k}.
\label{eq:app-ieee30-balancing-authority}
\end{equation}
For a positive deficit, eligible upward room is allocated proportionally and
then bounded emergency shedding is added if needed. Scheduled plus emergency
shedding at each bus is capped by 25\% of the balancing-load vector. For a
surplus, eligible generation is reduced first and scheduled shedding is then
cancelled proportionally. A remaining deficit or surplus is reported as
\textsf{Unresolved-deficit} or \textsf{Unresolved-surplus}. This procedure
preserves the authority boundary after local mode enforcement.

\subsection{Deployability and Hard Admissibility}

The implementation uses two causally ordered deployable sets.  The telemetry
set \(\mathcal{A}^{z}_{k}\) contains mode changes allowed by the telemetry
dwell, restriction, recovery, and emergency-isolation rules.  Each evaluated
telemetry candidate must satisfy the first three gates in
Eq.~\eqref{eq:app-ieee30-estimate-validity}; actual bad-data-test acceptance is evaluated
only after the selected current sample is acquired.  At the command stage,
\(\mathcal{A}^{u}_{k}\) is the complete admissible neighborhood through
Hamming distance two from \(\boldsymbol{\sigma}^{u}_{k-1}\), after command
dwell and transition rules.  For six generators and up to three alternative
modes per generator, it contains at most
\(1+6\cdot3+\binom{6}{2}3^{2}=154\) vectors.

For every \(\boldsymbol{\sigma}^{u}\in\mathcal{A}^{u}_{k}\), the predictor
uses the same accepted/fallback estimate, initial generation, forecast load,
dispatch request, and counterfactual reference.  The static hard screen is
\begin{equation}
\begin{aligned}
\boldsymbol{\sigma}^{u}\in\mathcal{F}^{u,\mathrm{stat}}_{k}
\Longleftrightarrow{}&
\text{dispatch succeeds},\\
&\mathbf{p}^{\min}\leq\mathbf{p}^{\mathrm{app}}
 \leq\mathbf{p}^{\max},\\
&\mathbf{0}\leq\mathbf{s}\leq\mathbf{d},
\quad
\mathbf{1}^{\mathsf T}\mathbf{s}
\leq0.10\,\mathbf{1}^{\mathsf T}\mathbf{d},\\
&|b_k|\leq10^{-6}\ \mathrm{MW},
\quad
|f_{\ell,k}|\leq1.15F_{\ell}\quad\forall\ell.
\end{aligned}
\label{eq:app-ieee30-static-hard-admissibility}
\end{equation}
In \textit{Dynamic and Timing-Aware}, a candidate must additionally avoid the dynamic hard conditions at
every planning quadrature node.  Normal-rating flow exceedance and crossings of
the 0.5-Hz or 0.5-Hz/s reference scales remain finite stress diagnostics and do
not make a candidate infeasible.  Response-aware and response-blind ablations
use the same candidate set and hard screen.

If no command candidate passes the hard screen, the search returns
\textsf{No-feasible-candidate} with reason codes and holds the preceding
mode vector.  The held vector is a declared fallback action, not an assertion
that the infeasible incumbent became admissible.  Predicted hard-screen events,
realized physical violations, estimator regimes, and undefined statistics are
logged separately.

\subsection{Objective-Term Construction and Normalization}

The staged objectives instantiate the four conceptual terms without pretending
to solve one global joint problem. For telemetry configuration
\(\boldsymbol{\sigma}^{z}\), the one-step propagation component is
\begin{equation}
R_{\mathrm{spr},k}(\boldsymbol{\sigma}^{z})
=\frac{1}{55}\sum_{j=1}^{55}
\widetilde p^{z}_{j,k+1}(\boldsymbol{\sigma}^{z}),
\label{eq:app-ieee30-spread-risk}
\end{equation}
where \(\widetilde p^{z}_{j,k+1}\) is given by
Eq.~\eqref{eq:app-ieee30-sis-belief-prediction}.

Let \(\mathfrak{C}_{k}=\{\mathbf{c}^{(m)}_{k}\}_{m=1}^{N_A}\) be the fixed
bank of nonzero state-displacement directions used for candidate comparison,
with corresponding state-consistent measurement perturbations
\(\mathbf{a}^{(m)}_{k}=\mathbf{H}\mathbf{c}^{(m)}_{k}\). Let
\(\mathbf{a}^{(m)}_{J,k}(\boldsymbol{\sigma}^{z})\) contain the rows of
attack \(m\) admitted by the candidate estimator. Its induced state
displacement and mean residual shift are
\begin{equation}
\begin{aligned}
\Delta\mathbf{x}^{(m)}_{k}(\boldsymbol{\sigma}^{z})
&=\mathbf{A}_{J,k}(\boldsymbol{\sigma}^{z})
  \mathbf{a}^{(m)}_{J,k}(\boldsymbol{\sigma}^{z}),\\
\boldsymbol{\mu}^{(m)}_{r,k}(\boldsymbol{\sigma}^{z})
&=\mathbf{Q}_{J,k}(\boldsymbol{\sigma}^{z})
  \mathbf{a}^{(m)}_{k}.
\end{aligned}
\label{eq:app-ieee30-fdi-effects}
\end{equation}
The retained displacement is normalized by the intended state displacement
\(\|\mathbf{c}^{(m)}_{k}\|_2\). Define
\begin{equation}
\begin{aligned}
d^{(m)}_{k}(\boldsymbol{\sigma}^{z})
&=\operatorname{clip}_{[0,1]}\!\left(
\frac{\|\Delta\mathbf{x}^{(m)}_{k}(\boldsymbol{\sigma}^{z})\|_{2}}
     {\|\mathbf{c}^{(m)}_{k}\|_{2}}\right),\\
\Lambda^{(m)}_{k}(\boldsymbol{\sigma}^{z})
&=\bigl(\boldsymbol{\mu}^{(m)}_{r,k}\bigr)^{\mathsf T}
  \mathbf{S}^{\dagger}_{r,k}
  \boldsymbol{\mu}^{(m)}_{r,k},\\
\pi^{(m)}_{\mathrm{byp},k}(\boldsymbol{\sigma}^{z})
&=F_{\chi^{2}_{\nu_k}(\Lambda^{(m)}_{k})}(\tau_k),\\
R_{\mathrm{FDI},k}(\boldsymbol{\sigma}^{z})
&=\frac{1}{N_A}\sum_{m=1}^{N_A}
d^{(m)}_{k}(\boldsymbol{\sigma}^{z})
\pi^{(m)}_{\mathrm{byp},k}(\boldsymbol{\sigma}^{z}).
\end{aligned}
\label{eq:app-ieee30-fdi-risk}
\end{equation}
Here, \(F_{\chi^{2}_{\nu}(\Lambda)}\) is the noncentral chi-square
cumulative distribution function, so
\(\pi^{(m)}_{\mathrm{byp},k}\) is the probability that the residual statistic
does not exceed the threshold \(\tau_k\). Candidates that fail the estimator
validity gates are excluded before this score is used.

The raw-ingress component is
\begin{equation}
R_{\mathrm{raw},k}(\boldsymbol{\sigma}^{z})
=\frac{1}{55}\sum_{j=1}^{55}
p^{z}_{j,k}v^{z}_{j}(\boldsymbol{\sigma}^{z}).
\label{eq:app-ieee30-raw-ingress-risk}
\end{equation}
The telemetry residual cyber risk, operational cost, and staged objective are
therefore
\begin{equation}
\begin{aligned}
R^{z}_{\mathrm{cyb},k}
&=\alpha_{\mathrm{spr}}R_{\mathrm{spr},k}
 +\alpha_{\mathrm{FDI}}R_{\mathrm{FDI},k}
 +\alpha_{\mathrm{raw}}R_{\mathrm{raw},k},\\
C^{z}_{\mathrm{op},k}
&=\beta_v C_{\mathrm{vis},k}+\beta_I C_{\mathrm{info},k},\\
J^{z}_{k}
&=\lambda^{z}_{\mathrm{cyb}}R^{z}_{\mathrm{cyb},k}
 +\lambda^{z}_{\mathrm{op}}C^{z}_{\mathrm{op},k}
 +\lambda^{z}_{\mathrm{sw}}C^{z}_{\mathrm{sw},k}.
\end{aligned}
\label{eq:app-ieee30-telemetry-objective}
\end{equation}
All component and outer weights are nonnegative and fixed before candidate
comparison; their reported values are given in the frozen parameterization.

The operational components are
\begin{equation}
\begin{aligned}
C_{\mathrm{vis},k}(\boldsymbol{\sigma}^{z})
&=1-\frac{1}{55}\sum_{j=1}^{55}
v^{z}_{j}(\boldsymbol{\sigma}^{z}),\\
C_{\mathrm{info},k}(\boldsymbol{\sigma}^{z})
&=\operatorname{clip}_{[0,1]}\!\left(
\frac{\rho_k(\boldsymbol{\sigma}^{z})-1}{\rho_{\max}-1}\right),\\
\rho_k(\boldsymbol{\sigma}^{z})
&=\frac{\operatorname{tr}(\mathbf{P}^{\mathrm{info}}_{J,k}
(\boldsymbol{\sigma}^{z}))}
   {\operatorname{tr}(\mathbf{P}^{\mathrm{info}}_{\textsf{Full},k})},
\end{aligned}
\label{eq:app-ieee30-telemetry-operational-components}
\end{equation}
where \(\rho_{\max}>1\) is the fixed information-quality gate and
\(\mathbf{P}^{\mathrm{info}}_{\textsf{Full},k}\) is the inverse information
matrix under the all-\textsf{Full} telemetry configuration. These quantities
expose a trade-off: raw visibility, structural observability, and the
information-quality proxy are related but not interchangeable.

After current-sample validation and construction of the single request
\(\mathcal{D}_k\), the command stage minimizes, over feasible local candidates,
\begin{equation}
J^{u}_{k}
=\lambda_z J^{z}_{k}
+\lambda_u R^{u}_{\mathrm{cyb},k}
+\lambda_{\mathrm{resp}}R_{\mathrm{resp},k}
+\lambda^{u}_{\mathrm{sw}}C^{u}_{\mathrm{sw},k}.
\label{eq:app-ieee30-command-objective}
\end{equation}
Although \(J^{z}_{k}\) is constant across command candidates, retaining it in
the reported score makes the staged decomposition explicit. Command risk is
the compromise-weighted retained central authority in
Eq.~\eqref{eq:ieee30-command-cyber-risk}. Both switching costs use normalized
ordinal mode distance, and transition/dwell constraints are applied before
candidate scoring.

For command candidate \(\boldsymbol{\sigma}^{u}\), let
\(\mathbf{f}^{\boldsymbol{\sigma}^{u}}_{k}\),
\(\mathbf{s}^{\boldsymbol{\sigma}^{u}}_{k}\), and
\(\mathbf{p}^{\mathrm{app},\boldsymbol{\sigma}^{u}}_{k}\) denote the
predicted branch flows, total modeled shedding, and applied generation after
mode enforcement and balancing. Define the unresolved active-power balance as
\begin{equation}
b^{\boldsymbol{\sigma}^{u}}_{k}
=\mathbf{1}^{\mathsf T}\!\left(
\mathbf{G}\mathbf{p}^{\mathrm{app},\boldsymbol{\sigma}^{u}}_{k}
-\mathbf{d}_{k}
+\mathbf{s}^{\boldsymbol{\sigma}^{u}}_{k}
\right),
\label{eq:app-ieee30-unresolved-balance}
\end{equation}
and let \(P^{\mathrm{base}}_{d}=\mathbf{1}^{\mathsf T}
\mathbf{d}^{\mathrm{base}}>0\). The three normalized static components are
\begin{equation}
\begin{aligned}
C_{f,k}(\boldsymbol{\sigma}^{u})
&=\frac{\sum_{\ell\in\mathcal{L}}
\left[|f^{\boldsymbol{\sigma}^{u}}_{\ell,k}|-F_{\ell}\right]_{+}}
{\sum_{\ell\in\mathcal{L}}F_{\ell}},\\
C_{s,k}(\boldsymbol{\sigma}^{u})
&=\frac{\mathbf{1}^{\mathsf T}
\mathbf{s}^{\boldsymbol{\sigma}^{u}}_{k}}
{P^{\mathrm{base}}_{d}},\\
C_{b,k}(\boldsymbol{\sigma}^{u})
&=\frac{|b^{\boldsymbol{\sigma}^{u}}_{k}|}
{P^{\mathrm{base}}_{d}}.
\end{aligned}
\label{eq:app-ieee30-static-components}
\end{equation}
Consequently,
let \(\mathcal{H}^{\mathrm{stat}}_{k}(\boldsymbol{\sigma}^{u})\) denote the
event that \(\boldsymbol{\sigma}^{u}\) violates at least one static condition
in Eq.~\eqref{eq:app-ieee30-static-hard-admissibility}. Then
\begin{equation}
\begin{aligned}
C^{\mathrm{stat}}_{k}(\boldsymbol{\sigma}^{u})
= {}&
\gamma_f C_{f,k}(\boldsymbol{\sigma}^{u})
+\gamma_s C_{s,k}(\boldsymbol{\sigma}^{u})
+\gamma_b C_{b,k}(\boldsymbol{\sigma}^{u})\\
&+\gamma_h\mathbf{1}_{\mathcal{H}^{\mathrm{stat}}_{k}(\boldsymbol{\sigma}^{u})},\\
R_{\mathrm{stat},k}(\boldsymbol{\sigma}^{u})
= {}&\left[
C^{\mathrm{stat}}_{k}(\boldsymbol{\sigma}^{u})
-C^{\mathrm{stat}}_{k}(\boldsymbol{\sigma}^{u,\mathrm{F}})
\right]_{+}.
\end{aligned}
\label{eq:app-ieee30-static-response-accounting}
\end{equation}
The hard indicator equals one when this event occurs and zero otherwise. The
positive part prevents a physical improvement relative to the
all-\textsf{Full} reference from becoming a negative risk credit; it does not
erase that improvement from the separately reported physical trajectory. In
\textit{Dynamic and Timing-Aware}, \(R_{\mathrm{resp},k}=R_{\mathrm{stat},k}+R_{\mathrm{dyn},k}\), with
the dynamic construction defined next.

\subsection{Independent Short-Horizon Dynamic Prediction}

\textit{Dynamic and Timing-Aware} appends a minimal aggregate
active-power/frequency layer to the quasi-static model. It is not a
multi-machine differential-algebraic, voltage, or transient line-flow model.
Let \(\Delta f(t)\) be center-of-inertia frequency deviation, \(p_m(t)\)
aggregate mechanical power, and \(p_e(t)\) served electrical demand. The
implemented dynamics are
\begin{equation}
\begin{aligned}
M_f\,\dot{\Delta f}(t)
&=p_m(t)-p_e(t)-D\Delta f(t),
\qquad
M_f=\frac{2H_{\mathrm{eq}}S_M}{f_0},\\
T_g\dot p_m(t)
&=p_{\mathrm{ref}}(t)-p_m(t)-K_R\Delta f(t).
\end{aligned}
\label{eq:app-ieee30-dynamic-model}
\end{equation}
For mode-dependent activation time \(\delta_g\),
\begin{equation}
p_{\mathrm{ref}}(t)
=\sum_{g\in\mathcal{G}}
\begin{cases}
p^{0}_{g,k},&t<\delta_g,\\
p^{\mathrm{app}}_{g,k},&t\geq\delta_g,
\end{cases}
\label{eq:app-ieee30-delayed-reference}
\end{equation}
where \(\delta_g=\infty\) for \textsf{Isolated}. Remote jitter is added only
to \textsf{Full} and \textsf{Guarded} delays; the \textsf{Safe-local} delay is
deterministic. Because the quasi-static \textsf{Safe-local} target equals the
held output, its activation does not imply an undeclared redispatch action.

Each epoch is an independent five-second timing-sensitivity snapshot:
\begin{equation}
\Delta f(0)=0,
\qquad
p_m(0)=\mathbf{1}^{\mathsf T}\mathbf{p}^{0}_{k}.
\label{eq:app-ieee30-dynamic-initialization}
\end{equation}
Candidate, all-\textsf{Full} reference, and realized branches share this
initialization. Terminal dynamic state is not carried to the next decision
epoch. The internal integration grid is the union of the 0.05-s reporting grid
and every finite command-activation time. The command reference valid at the
subinterval midpoint is used in a second-order predictor--corrector step, so an
event such as 0.02~s is not rounded or smeared over the preceding interval.

The selector does not observe future execution jitter. Instead, it uses the
frozen three-node rule
\begin{equation}
\begin{aligned}
z_q&\in\{-\sqrt{3},0,\sqrt{3}\},\\
w_q&\in\left\{\frac16,\frac23,\frac16\right\},\\
\epsilon_q&=\operatorname{clip}(\sigma_{\epsilon}z_q,
-\epsilon_{\max},\epsilon_{\max}),
\end{aligned}
\label{eq:app-ieee30-jitter-quadrature}
\end{equation}
where \(\sigma_{\epsilon}\) is the jitter standard deviation and
\(\epsilon_{\max}\) its clipping bound. Candidate and reference are evaluated
at the same node. The realized jitter is sampled only after selection from a
policy-independent common-random stream.

Let \(T\) be the horizon, \(f_s=0.5\)~Hz the frequency-deviation stress scale,
\(r_s=0.5\)~Hz/s the RoCoF stress scale, \(t_{\mathrm{set}}\) the settling
time within the 0.05-Hz band, and \(\Delta f(T)\) the terminal error. The
continuous predictive stress is
\begin{equation}
\begin{aligned}
S^{\mathrm{pred}}
=\frac14\Bigg(&
\frac{1}{T}\int_{0}^{T}
\left(\frac{|\Delta f(t)|}{f_s}\right)^2\!\mathrm{d}t
+\frac{1}{T}\int_{0}^{T}
\left(\frac{|\dot{\Delta f}(t)|}{r_s}\right)^2\!\mathrm{d}t\\
&+\frac{t_{\mathrm{set}}}{T}
+\left(\frac{|\Delta f(T)|}{f_s}\right)^2
\Bigg).
\end{aligned}
\label{eq:app-ieee30-predictive-stress}
\end{equation}
The dynamic decision severity is
\(S^{\mathrm{dyn}}=S^{\mathrm{pred}}+C^{\mathrm{dyn,hard}}\), where the hard
cost is zero unless a declared hard condition is crossed. Dynamic response
risk is the expected nodewise positive increment
\begin{equation}
R_{\mathrm{dyn},k}
=\sum_q w_q
\left[
S^{\mathrm{dyn}}_{k}(\boldsymbol{\sigma}^{u};\epsilon_q)
-S^{\mathrm{dyn}}_{k}(\boldsymbol{\sigma}^{u,\mathrm{F}};\epsilon_q)
\right]_{+}.
\label{eq:app-ieee30-dynamic-response-risk}
\end{equation}
Taking the positive part before the weighted expectation preserves the paired
candidate--reference comparison at each uncertainty node.

The 0.5-Hz and 0.5-Hz/s values in
Eq.~\eqref{eq:app-ieee30-predictive-stress} are reporting and normalization
scales, not certified protection limits. Let \(s^{\mathrm{UFLS}}\) be the
under-frequency load shedding and \(P_d\) the total demand. The dynamic hard
predicate is
\begin{equation}
\begin{aligned}
\mathcal{H}^{\mathrm{dyn}}
={}&\{\text{numerical instability}\}\\
&\cup\{\max_t|\Delta f(t)|>1.5~\mathrm{Hz}\}\\
&\cup\{s^{\mathrm{UFLS}}/P_d>0.10\}.
\end{aligned}
\label{eq:app-ieee30-dynamic-hard-predicate}
\end{equation}
No hard RoCoF predicate is declared. Under-frequency load shedding is
implemented but disabled in the reported primary campaign. Consequently,
\textit{Dynamic and Timing-Aware} supports a short-horizon timing and
decision-relevance analysis; it does not establish recursive dynamic safety or
multi-machine stability.


\section{Corrective Dispatch and Attack-Independent Calibration}
\label{app:ieee30-dispatch}

This appendix specifies the controller assumptions and calibration procedure
used in Section~\ref{sec:ieee30-experimental-evaluation}.  They are fixed before
the paired policy comparisons.

\textbf{Finite operating envelopes.}
The six generator bounds recovered from \texttt{case\_ieee30} are
\begin{equation}
\begin{aligned}
\mathbf{p}^{\min}_{g}
&=[0,0,0,0,0,0]~\mathrm{MW},\\
\mathbf{p}^{\max}_{g}
&=[360.2,140,100,100,100,100]~\mathrm{MW}.
\end{aligned}
\label{eq:app-ieee30-generator-limits}
\end{equation}
Because the source case provides no finite generator ramp data, the
quasi-static per-epoch flexibility is declared as
\begin{equation}
\Delta p^{\max}_{g}
=\max\!\left\{
0.20(p^{\max}_{g}-p^{\min}_{g}),
5~\mathrm{MW}
\right\}.
\label{eq:app-ieee30-flexibility}
\end{equation}
This is a decision-epoch envelope, not a generator-specific MW/s or MW/min
rating.  The 41 normal branch envelopes use the 16--130 MVA
\texttt{RATE\_A} pattern from the companion MATPOWER \texttt{case30}, mapped
by branch endpoints to the common topology.  Within the DC active-power model,
the scaled numerical values act as benchmark MW flow envelopes rather than
validated alternating-current thermal ratings.

\textbf{Forecast, realized load, and spatial inference.}
At epoch \(k\), the selector receives
\begin{equation}
\mathbf{d}^{\mathrm{fc}}_{k}
=\mathbf{d}^{\mathrm{base}}h_kL,
\qquad
P^{\mathrm{agg}}_{d,k}
=\mathbf{1}^{\mathsf T}\mathbf{d}^{\mathrm{fc}}_{k},
\label{eq:app-ieee30-load-forecast}
\end{equation}
where \(h_k\) is the public profile multiplier and \(L\) the declared load
stress.  The episode-specific stochastic multiplier is applied afterward to
form \(\mathbf{d}^{\mathrm{real}}_{k}\).  The selector never receives this
realized vector.  Actuator balancing receives only the protected aggregate
mismatch
\begin{equation}
\Delta P^{\mathrm{prot}}_{d,k}
=\mathbf{1}^{\mathsf T}\mathbf{d}^{\mathrm{real}}_{k}
-\mathbf{1}^{\mathsf T}\mathbf{d}^{\mathrm{fc}}_{k}.
\label{eq:app-ieee30-protected-mismatch}
\end{equation}

Given accepted or permitted fallback estimate \(\widetilde{\mathbf{x}}_k\)
and current generation \(\mathbf{p}^{0}_{k}\), the controller forms the raw
load pattern
\begin{equation}
\mathbf{d}^{\mathrm{raw}}_{k}
=\mathbf{G}\mathbf{p}^{0}_{k}
-S_B\mathbf{B}_{\mathrm{bus}}\mathbf{T}_{r}
\widetilde{\mathbf{x}}_k,
\quad
\mathbf{B}_{\mathrm{bus}}=\mathbf{C}^{\mathsf T}
\mathbf{B}_{\ell}\mathbf{C}.
\label{eq:app-ieee30-raw-load-inference}
\end{equation}
Each component is clipped to
\([0,4\max\{d^{\mathrm{base}}_n,1~\mathrm{MW}\}]\).  If the clipped vector
has zero total, the base-load pattern is used.  The resulting nonnegative
pattern is rescaled so that
\begin{equation}
\mathbf{1}^{\mathsf T}\widehat{\mathbf{d}}_{k}
=P^{\mathrm{agg}}_{d,k}.
\label{eq:app-ieee30-fixed-demand}
\end{equation}
An \ac{fdia} can therefore alter the inferred spatial distribution and the
resulting redispatch, but cannot create aggregate forecast demand.

\textbf{Corrective dispatch.}
Together with the spatial-inference construction in
Eqs.~\eqref{eq:app-ieee30-raw-load-inference}--
\eqref{eq:app-ieee30-fixed-demand}, the following optimization defines the
corrective-dispatch mapping \(\mathcal{K}\) in
Eq.~\eqref{eq:ieee30-dispatch-request}. Its context
\(\boldsymbol{\xi}_{k}\) supplies the fixed operating limits, controller
headroom, and other controller parameters available at decision time.
Let \(\mathbf{p}_{g}\) be requested generation, \(\mathbf{s}\) scheduled
shedding, and \(\boldsymbol{\delta}^{+},\boldsymbol{\delta}^{-}\) positive and
negative redispatch auxiliaries. The controller solves
\begin{equation}
\begin{aligned}
\min_{\mathbf{p}_{g},\mathbf{s},
      \boldsymbol{\delta}^{+},\boldsymbol{\delta}^{-}}
\quad&
\mathbf{c}_{g}^{\mathsf T}\mathbf{p}_{g}
+c_{\Delta}\mathbf{1}^{\mathsf T}
(\boldsymbol{\delta}^{+}+\boldsymbol{\delta}^{-})
+c_{s}\mathbf{1}^{\mathsf T}\mathbf{s}
\end{aligned}
\label{eq:app-ieee30-dispatch-objective}
\end{equation}
subject to
\begin{equation}
\begin{aligned}
\mathbf{1}^{\mathsf T}\mathbf{p}_{g}
+\mathbf{1}^{\mathsf T}\mathbf{s}
&=\mathbf{1}^{\mathsf T}\widehat{\mathbf{d}}_{k},\\
\mathbf{p}^{\min}_{g}
\leq\mathbf{p}_{g}
&\leq\mathbf{p}^{\max}_{g},\\
\mathbf{p}_{g}-\mathbf{p}^{0}_{g,k}
&=\boldsymbol{\delta}^{+}-\boldsymbol{\delta}^{-},\\
|\mathbf{p}_{g}-\mathbf{p}^{0}_{g,k}|
&\leq\Delta\mathbf{p}^{\max}_{g},\\
\boldsymbol{\delta}^{+},\boldsymbol{\delta}^{-}
&\geq\mathbf{0},\\
|\widehat{\mathbf{f}}(\mathbf{p}_{g},\mathbf{s})|
&\leq\mathbf{F}^{\mathrm{ctrl}},\\
\mathbf{0}\leq\mathbf{s}
&\leq0.20\widehat{\mathbf{d}}_{k}.
\end{aligned}
\label{eq:app-ieee30-dispatch-constraints}
\end{equation}
If the problem is feasible with optimizer
\((\mathbf{p}^{\star}_{g,k},\mathbf{s}^{\star}_{k})\), then
\begin{equation}
\mathbf{p}^{\mathrm{req}}_{k}=\mathbf{p}^{\star}_{g,k},
\qquad
\mathbf{s}^{\mathrm{req}}_{k}=\mathbf{s}^{\star}_{k},
\qquad
\widehat{\mathbf{f}}_{k}
=\widehat{\mathbf{f}}(\mathbf{p}^{\star}_{g,k},\mathbf{s}^{\star}_{k}),
\,
\mathrm{status}_{k}=\textsf{Success}.
\label{eq:app-ieee30-dispatch-outputs}
\end{equation}
The shedding penalty \(c_s\) is much larger than the generation and redispatch
costs. A failed solve sets \(\mathrm{status}_{k}=\textsf{Failure}\) and
returns a diagnostic hold request; it is hard-rejected during ordinary command
candidate screening. The same successful \(\mathcal{D}_k\) is used in
candidate scoring, actual enforcement, logging, and the all-\textsf{Full}
reference.

\input{Tables/Evaluation_params}

\textbf{Controller headroom.}
For attack-free calibration sample \(s\), define the positive branch-flow
underprediction error
\begin{equation}
e_{\ell,s}
=\left[
|f^{\mathrm{actual}}_{\ell,s}|
-|f^{\mathrm{pred}}_{\ell,s}|
\right]_{+},
\qquad
e^{\max}_{s}=\max_{\ell}e_{\ell,s}.
\label{eq:app-ieee30-headroom-errors}
\end{equation}
With the empirical higher-quantile operator \(Q_{0.99}\), the implemented
margin is
\begin{equation}
m_{\ell}
=\max\!\left\{
Q_{0.99}(\{e_{\ell,s}\}_{s=1}^{512}),
Q_{0.99}(\{e^{\max}_{s}\}_{s=1}^{512})
\right\}.
\label{eq:app-ieee30-headroom}
\end{equation}
Because \(e_{\ell,s}\leq e^{\max}_{s}\) for every sample, the system term
dominates under the common empirical quantile convention; the retained design
therefore acts as a common absolute MW margin across branches.  An independent
256-sample run validates the frozen margin without changing it.  Physical
overload remains defined against \(F_{\ell}\), not
\(F^{\mathrm{ctrl}}_{\ell}\).

\textbf{Clean stress selection.}
For each predeclared load and branch-scale pair, demand is scaled, reference
generation is allocated by the bounded proportional rule, and the attack-free
DC state is solved before corrective dispatch.  Only points satisfying clean
dispatch, balance, shedding, headroom, and hard feasibility remain eligible.
The eligible point closest to the predeclared loading target 0.75 defines the
reference capacity scale; attack outcomes and relative policy performance do
not enter this selection.

\section{Frozen Evaluation Parameters}
\label{app:ieee30-campaign-parameters}

This appendix reports the parameters fixed for the comparisons in
Section~\ref{sec:ieee30-experimental-evaluation}.  They define an illustrative
scenario bank and policy priorities; they are not fitted estimates of an
operational transmission system.

\textbf{Campaign design.}
Table~\ref{tab:app-ieee30-campaign-design} summarizes the paired scenario bank
for each evaluation configuration. Within a comparison, every evaluated policy receives the same
compromise, alert, attack, measurement-noise, load, and disturbance
realizations. \textit{Dynamic and Timing-Aware} additionally shares the
post-selection realization of authorization jitter.

The hidden compromise process starts with five uniformly selected telemetry
endpoints.  At each epoch, a susceptible endpoint is infected from each
compromised neighbor with probability \(\beta=0.1\), whereas a compromised
endpoint recovers with probability \(\gamma=0.2\).  Defender-visible alerts
have detection probability 0.90 and false-alarm probability 0.05.  Measurement
noise uses relative standard deviation 0.02 with a 0.001-p.u. floor, and the
bad-data detector uses significance level 0.05.

\textbf{Telemetry-policy parameters.}
The \textit{Graduated Telemetry} residual cyber-risk term and telemetry objective are
\begin{equation}
R^{z}_{\mathrm{cyb},k}
=0.25R_{\mathrm{spr},k}
+0.50R_{\mathrm{FDI},k}
+0.25R_{\mathrm{raw},k},
\label{eq:app-ieee30-stage1-cyber-objective}
\end{equation}
\begin{equation}
J^{z}_{k}
=0.60R^{z}_{\mathrm{cyb},k}
+0.30C^{z}_{\mathrm{op},k}
+0.10C^{z}_{\mathrm{sw},k}.
\label{eq:app-ieee30-stage1-objective}
\end{equation}
Within \(C^{z}_{\mathrm{op},k}\), visibility loss and information-proxy
degradation receive weights 0.30 and 0.70.  \textsf{Guarded} retains peer
authority 0.25 and relative estimator precision 0.35.  The maximum
inverse-information trace ratio is \(\rho_{\max}=5\), and mandatory raw-stream
isolation applies at compromise belief 0.90.  Evidence-dependent broker
refinement is disabled, so the reported campaign attributes no additional
benefit to within-mode adaptive precision.  The telemetry heuristic considers
at most ten suspicious/recovering endpoints and accepts at most four improving
mode changes per epoch; it is not described as a global optimizer.

\textbf{Command-policy parameters.}
For \textit{Static Consequence-Aware}, the response-aware selector minimizes
\begin{equation}
J^{u,\mathrm{stat}}_{k}
=0.45J^{z}_{k}
+0.20R^{u}_{\mathrm{cyb},k}
+0.25R_{\mathrm{stat},k}
+0.10C^{u}_{\mathrm{sw},k}.
\label{eq:app-ieee30-stage2-objective}
\end{equation}
The response-blind ablation removes only \(R_{\mathrm{stat},k}\) and
renormalizes the remaining coefficients while preserving their ratios:
\begin{equation}
J^{u,\mathrm{blind}}_{k}
=0.60J^{z}_{k}
+0.266\overline{6}R^{u}_{\mathrm{cyb},k}
+0.133\overline{3}C^{u}_{\mathrm{sw},k}.
\label{eq:app-ieee30-stage2-blind-objective}
\end{equation}
\textit{Dynamic and Timing-Aware} retains the \textit{Static Consequence-Aware} weights and replaces \(R_{\mathrm{stat},k}\) by
\(R_{\mathrm{stat},k}+R_{\mathrm{dyn},k}\).  Command restriction, recovery,
and emergency thresholds are 0.20, 0.15, and 0.95, respectively, with one
epoch of minimum dwell.  The selector enumerates the entire allowed
Hamming-distance-two neighborhood and applies deterministic tie-breaking after the
common hard screen.  \textsf{Guarded} retains 35\% of ordinary command
flexibility and central-command authority.  Static consequence weights are
1 for normalized normal-envelope overload, 5 for normalized shedding, 5 for
normalized unresolved balance, and 1 for a hard-infeasibility indicator.  The
hard branch ceiling is 1.15 times the normal envelope and the balance tolerance
is \(10^{-6}\)~MW.

\textbf{Dynamic and authorization parameters.}
The aggregate model uses nominal frequency \(f_0=60\)~Hz, machine base
\(S_M=451.2\)~MVA, equivalent inertia
\(H_{\mathrm{eq}}=3.45012\)~s, damping
\(D=10.70667\)~MW/Hz, droop gain
\(K_R=248.84622\)~MW/Hz, and governor time constant
\(T_g=0.26554\)~s.  The frequency-deviation and RoCoF normalization scales are
0.5~Hz and 0.5~Hz/s, the settling band is 0.05~Hz, and the synthetic hard
frequency-deviation ceiling is 1.5~Hz.  No hard RoCoF limit is declared.
Under-frequency load shedding is disabled in the reported primary campaigns;
if enabled, a UFLS fraction above 0.10 is a hard event.

Default activation delays are 0.10~s for \textsf{Full}, 0.20~s for
\textsf{Guarded}, and 0.02~s for \textsf{Safe-local}.  Remote jitter has
standard deviation 0.03~s and is clipped at \(\pm0.10\)~s.  Planning uses the
three quadrature nodes in Eq.~\eqref{eq:app-ieee30-jitter-quadrature}; execution
jitter is sampled after selection.  The paired slower-authorization
sensitivity changes only the \textsf{Full} and \textsf{Guarded} delays to
0.40~s and 0.60~s.  The local-fallback delay and all other model, policy, and
scenario parameters remain unchanged.

Controller-headroom calibration uses 512 attack-free samples with seed
20260809 and 256 independent validation samples.  The operating-stress family
and its selection rule are specified in Appendix~\ref{app:ieee30-dispatch}.

\section{Terminology}   \label{sec:Terminology}
This appendix consolidates the notation used in the domain-independent
formulation and the IEEE 30-bus instantiation.

\input{Tables/Terminology}

%% file: Tables/Evaluation_params.tex
\begin{table*}[!t]
\centering
\caption{Frozen campaign design for the three evaluation configurations.}
\label{tab:app-ieee30-campaign-design}
\small
\begin{tabularx}{\textwidth}{@{}p{0.14\textwidth}p{0.27\textwidth}p{0.25\textwidth}X@{}}
\toprule
Configuration & Scenario bank & Operating point & Additional parameters \\
\midrule
\textit{Graduated Telemetry}
& 2000 episodes, 40 epochs per episode, 1000 fixed attack directions, and fixed seed 20260727.
& IEEE 30-bus cyber and measurement model; five initially compromised telemetry endpoints.
& \ac{fdia} magnitude 0.5 times the measurement-noise norm; impactful-displacement threshold 0.05. \\
\addlinespace
\textit{Static Consequence-Aware}
& For each load multiplier, 200 episodes, 40 epochs, 500 fixed attack directions, and fixed seed 20260806.
& \(L\in\{1.00,1.05,1.10\}\) and branch-envelope scale \(S_F=1.70\).
& \textit{Graduated Telemetry} attack magnitude; load-disturbance standard deviation 0.02. \\
\addlinespace
\textit{Dynamic and Timing-Aware}
& 200 episodes, 40 epochs, 500 fixed attack directions, and fixed seed 20260806.
& \(L=1.10\) and \(S_F=1.70\).
& \textit{Static Consequence-Aware} settings; independent 5-s dynamic snapshots and 0.05-s reporting cadence with event-time substeps. \\
\bottomrule
\end{tabularx}
\end{table*}

%% file: Tables/Terminology.tex
\begin{table*}[!p]
\centering
\renewcommand{\arraystretch}{1.15}
\caption{Core notation used in the domain-independent \ac{sa-zt} formulation and its IEEE 30-bus instantiation.}
\label{tab:general-notation}
\scriptsize
\begin{tabular}{@{}p{0.30\textwidth}|p{0.66\textwidth}@{}}
\toprule
\textbf{Symbol} & \textbf{Meaning} \\
\midrule
\multicolumn{2}{@{}l}{\textit{Domain-independent \ac{sa-zt} formulation}}\\
\midrule

\(k\)
& Runtime decision epoch. \\

\(\mathcal{E}=\{1,\ldots,N\}\)
& Set of \(N\) independently enforceable units, indexed by \(i\). \\

\(\Sigma_i\)
& Set of enforcement modes supported by unit \(i\). \\

\(\boldsymbol{\sigma}=(\sigma_1,\ldots,\sigma_N)\)
& Candidate joint configuration, with \(\sigma_i\in\Sigma_i\). \\

\(\boldsymbol{\sigma}_{k}\), \(\boldsymbol{\sigma}^{\star}_{k}\)
& Configuration enforced at epoch \(k\), and configuration selected by the Safety Engine, respectively. \\

\(\mathcal{B}_k\)
& Uncertainty-aware cyber belief derived from the security evidence available at epoch \(k\). \\

\(\boldsymbol{\xi}_k\)
& Defender-observable operating context, including the estimated process state, operating mode, objectives, deadlines, and uncertainty. \\

\(\mathcal{A}_k\), \(\mathcal{F}_k\subseteq\mathcal{A}_k\)
& Configurations that are authorized, supported, and deployable, and the subset satisfying all declared hard operational and physical conditions. \\

\(\mathcal{D}(\boldsymbol{\sigma})
=(\mathcal{U}(\boldsymbol{\sigma}),
\mathcal{V}(\boldsymbol{\sigma}),
\mathcal{W}(\boldsymbol{\sigma}))\)
& Capability interface induced by \(\boldsymbol{\sigma}\): retained state-changing authority \(\mathcal{U}\), raw telemetry visibility \(\mathcal{V}\), and maximum automated influence \(\mathcal{W}\). \\

\(R_{\mathrm{cyb},k},C_{\mathrm{op},k},
R_{\mathrm{resp},k},C_{\mathrm{sw},k}\)
& Residual cyber risk, non-safety-critical degradation of legitimate operation, response-induced physical risk, and non-physical switching or recovery cost. \\

\(C_{\mathrm{phys},k}^{\boldsymbol{\sigma},0}\),
\(C_{\mathrm{phys},k}^{\mathrm{ref},0}\)
& Predicted physical consequence under the candidate and fixed reference responses, respectively, with no additional adversarial action introduced after epoch \(k\). \\

\(\Delta C_{\mathrm{phys},k}^{\boldsymbol{\sigma},0}\)
& Incremental physical consequence
\(C_{\mathrm{phys},k}^{\boldsymbol{\sigma},0}
-C_{\mathrm{phys},k}^{\mathrm{ref},0}\);
\((a)_{+}=\max\{a,0\}\) retains only positive increments when computing response-induced risk. \\

\(\lambda_{\mathrm{cyb}},\lambda_{\mathrm{op}},
\lambda_{\mathrm{resp}},\lambda_{\mathrm{sw}}\)
& Nonnegative coefficients representing declared policy priorities. \\

\midrule
\multicolumn{2}{@{}l}{\textit{IEEE 30-bus instantiation}}\\
\midrule

\(\mathcal{N},\mathcal{L},\mathcal{G},\mathcal{N}^{p}\)
& Bus set, branch set, generator-bus set, and subset of buses with injection measurements, respectively. \\

\(z,u,f,p\)
& Superscripts identifying telemetry-side, command-side, branch-flow, and bus-injection quantities, respectively. \\

\(\mathcal{E}^{z},\mathcal{E}^{u}\)
& Sets of telemetry and generator-command enforcement units, respectively. \\

\(e^{f}_{\ell},e^{p}_{n},e^{u}_{g}\)
& Logical enforcement units associated with branch-flow measurement \(\ell\), injection measurement \(n\), and generator-command interface \(g\), respectively. \\

\(\boldsymbol{\sigma}^{z}_{k},\boldsymbol{\sigma}^{u}_{k},
\boldsymbol{\sigma}^{u,\mathrm{F}}\)
& Telemetry and command configurations at epoch \(k\), and the all-\textsf{Full} command reference configuration. \\

\(\mathbf{x}_{k}\)
& Reduced DC state vector containing the 29 non-reference bus-voltage angles. \\

\(\mathbf{H}^{f},\mathbf{H}^{p},\mathbf{H}\)
& Branch-flow measurement block, bus-injection measurement block, and complete measurement matrix. \\

\(\boldsymbol{\Pi}_{z}\)
& Fixed permutation placing the stacked measurements in logical telemetry-endpoint order. \\

\(\mathbf{z}^{\mathrm{raw}}_{k}\)
& Raw telemetry vector received before policy-mediated estimation. \\

\(\boldsymbol{\eta}_{k},\mathbf{R}_{k},\mathbf{a}_{k}\)
& Physical measurement noise, its covariance, and the attack realized through compromised telemetry endpoints. \\

\(\mathbf{v}^{z}(\boldsymbol{\sigma}^{z}),
\boldsymbol{\omega}^{z}(\boldsymbol{\sigma}^{z})\)
& Raw-visibility vector and relative \ac{wls}-precision vector induced by a telemetry configuration. \\

\(\widetilde{\mathbf{x}}_{k},\mathbf{p}^{0}_{k},
P^{\mathrm{agg}}_{d,k}\)
& Accepted current state estimate or permitted one-epoch stale estimate, current validated generator-output vector, and protected aggregate demand forecast. \\

\(\mathcal{K},\mathcal{D}_{k}\)
& Corrective-dispatch mapping and the single dispatch request produced for epoch \(k\). \\

\(\mathbf{p}^{z}_{k},\mathbf{p}^{u}_{k}\)
& Estimated compromise-probability vectors for telemetry and command interfaces; together they instantiate \(\mathcal{B}_{k}\). \\

\(R^{z}_{\mathrm{cyb},k}\)
& Telemetry residual cyber risk. \\

\(R_{\mathrm{spr},k},R_{\mathrm{FDI},k},R_{\mathrm{raw},k}\)
& Predicted compromise propagation, retained \ac{fdia} effect weighted by bad-data bypass probability, and compromise-weighted raw-ingress exposure. \\

\(\alpha_{\mathrm{spr}},\alpha_{\mathrm{FDI}},\alpha_{\mathrm{raw}}\)
& Nonnegative weights of the telemetry cyber-risk components. \\

\(C^{z}_{\mathrm{op},k},
C_{\mathrm{vis},k},C_{\mathrm{info},k}\)
& Telemetry operational cost and its raw-visibility-loss and information-degradation components. \\

\(\beta_v,\beta_I\)
& Nonnegative weights of the telemetry operational-cost components. \\

\(R^{u}_{\mathrm{cyb},k},p^{u}_{g,k},
a_g(\sigma^{u}_{g})\)
& Retained command cyber risk, compromise probability of command interface \(g\), and central-command authority retained by its selected mode. \\

\(C^{\mathrm{stat}}_{k},
C_{f,k},C_{s,k},C_{b,k}\)
& Static physical-consequence score and its normalized branch-exceedance, modeled-shedding, and unresolved-balance components. \\

\(\mathbb{1}_{\mathcal{H}^{\mathrm{stat}}_k}
(\boldsymbol{\sigma}^{u})\)
& Indicator equal to one when the candidate command configuration violates at least one declared static hard condition, and zero otherwise. \\

\(\gamma_f,\gamma_s,\gamma_b,\gamma_h\)
& Nonnegative weights applied to the branch-exceedance, modeled-shedding, unresolved-balance, and static-hard-violation terms, respectively. \\

\(F_{\ell}\)
& Normal operating-envelope rating of branch \(\ell\); \(1.15F_{\ell}\) is the synthetic hard ceiling used in the case study. \\

\(R_{\mathrm{stat},k},R_{\mathrm{dyn},k}\)
& Static and short-horizon dynamic components of response-induced physical risk. \\

\(C^{z}_{\mathrm{sw},k},C^{u}_{\mathrm{sw},k}\)
& Normalized telemetry- and command-mode switching costs. \\

\bottomrule
\end{tabular}
\end{table*}

%% file: Bibliography.bib
@ARTICLE{CP-ZTA,
  author={Feng, Xiaomeng and Hu, Shiyan},
  journal={IEEE Transactions on Industrial Cyber-Physical Systems}, 
  title={Cyber-Physical Zero Trust Architecture for Industrial Cyber-Physical Systems}, 
  year={2023},
  volume={1},
  number={},
  pages={394-405},
  doi={10.1109/TICPS.2023.3333850}
}

@unpublished{Converging_ZT,
  TITLE = {{Converging Zero Trust and IoT Security: A Multivocal Literature Review}},
  AUTHOR = {Wehbe, Mariam and Bobelin, Laurent},
  URL = {https://hal.science/hal-05577931},
  NOTE = {working paper or preprint},
  YEAR = {2025},
  MONTH = Sep,
  HAL_ID = {hal-05577931},
  HAL_VERSION = {v1}
}

@article{Verify_and_trust,
title = {Verify and trust: A multidimensional survey of zero-trust security in the age of IoT},
journal = {Internet of Things},
volume = {27},
pages = {101227},
year = {2024},
issn = {2542-6605},
doi = {https://doi.org/10.1016/j.iot.2024.101227},
url = {https://www.sciencedirect.com/science/article/pii/S2542660524001689},
author = {Muhammad Ajmal Azad and Sidrah Abdullah and Junaid Arshad and Harjinder Lallie and Yussuf Hassan Ahmed}
}

@article{Survey_ZT,
author = {He, Yuanhang and Huang, Daochao and Chen, Lei and Ni, Yi and Ma, Xiangjie},
title = {A Survey on Zero Trust Architecture: Challenges and Future Trends},
journal = {Wireless Communications and Mobile Computing},
volume = {2022},
number = {1},
pages = {6476274},
doi = {https://doi.org/10.1155/2022/6476274},
url = {https://onlinelibrary.wiley.com/doi/abs/10.1155/2022/6476274},
eprint = {https://onlinelibrary.wiley.com/doi/pdf/10.1155/2022/6476274},
year = {2022}
}

@INPROCEEDINGS{TrustAware_ZT,
  author={Dimitrakos, Theo and Dilshener, Tezcan and Kravtsov, Alexander and La Marra, Antonio and Martinelli, Fabio and Rizos, Athanasios and Rosetti, Alessandro and Saracino, Andrea},
  booktitle={2020 IEEE 19th International Conference on Trust, Security and Privacy in Computing and Communications (TrustCom)}, 
  title={Trust Aware Continuous Authorization for Zero Trust in Consumer Internet of Things}, 
  year={2020},
  volume={},
  number={},
  pages={1801-1812}
  }

@article{ComprehensiveReview_ZT,
  title={A comprehensive review and comparative analysis of zero trust architecture: Evolution, implementation strategies, and key challenges},
  author={Soni, Ashutosh and Kumar Nanda, Surendra and Priyadarshini, Rojalina and Panda, Ganapati},
  journal={Journal of Computer Security},
  volume={34},
  number={2},
  pages={85--110},
  year={2026},
  publisher={SAGE Publications Sage UK: London, England}
}

@article{SystematicReview_ZT,
  title={A systematic literature review on the implementation and challenges of zero trust architecture across domains},
  author={Sadaf, Mushtaq and Muhammad, Mohsin and Mujahid, Mushtaq Muhammad},
  journal={Sensors},
  volume={25},
  number={19},
  pages={6118},
  year={2025},
  publisher={MDPI AG}
}

@Article{ReviewComparative_ZT,
AUTHOR = {Dhiman, Poonam and Saini, Neha and Gulzar, Yonis and Turaev, Sherzod and Kaur, Amandeep and Nisa, Khair Ul and Hamid, Yasir},
TITLE = {A Review and Comparative Analysis of Relevant Approaches of Zero Trust Network Model},
JOURNAL = {Sensors},
VOLUME = {24},
YEAR = {2024},
NUMBER = {4},
ARTICLE-NUMBER = {1328},
URL = {https://www.mdpi.com/1424-8220/24/4/1328},
PubMedID = {38400486},
ISSN = {1424-8220},
DOI = {10.3390/s24041328}
}

@article{FutureIndustry_ZT,
  title={Future industry internet of things with zero-trust security},
  author={Li, Shan and Iqbal, Muddesar and Saxena, Neetesh},
  journal={Information Systems Frontiers},
  volume={26},
  number={5},
  pages={1653--1666},
  year={2024},
  publisher={Springer}
}

@Article{Theory_ZT,
AUTHOR = {Kang, Hongzhaoning and Liu, Gang and Wang, Quan and Meng, Lei and Liu, Jing},
TITLE = {Theory and Application of Zero Trust Security: A Brief Survey},
JOURNAL = {Entropy},
VOLUME = {25},
YEAR = {2023},
NUMBER = {12},
ARTICLE-NUMBER = {1595},
URL = {https://www.mdpi.com/1099-4300/25/12/1595},
PubMedID = {38136475},
ISSN = {1099-4300},
DOI = {10.3390/e25121595}
}

@article{ZT_ContextIoT,
  title={Zero Trust in the Context of IoT: Industrial Literature Review, Trends, and Challenges.},
  author={Bobelin, Laurent},
  journal={C\&ESAR},
  pages={37--52},
  year={2023}
}

@article{Critical_ZT,
  title={A critical analysis of Zero Trust Architecture ({ZTA})},
  author={Fernandez, Eduardo Buglioni and Brazhuk, Andrei},
  journal={Available at SSRN 4210104},
  year={2022}
}

@article{Towards_ZT,
title = {Towards zero trust security in connected vehicles: A comprehensive survey},
journal = {Computers \& Security},
volume = {145},
pages = {104018},
year = {2024},
issn = {0167-4048},
doi = {https://doi.org/10.1016/j.cose.2024.104018},
url = {https://www.sciencedirect.com/science/article/pii/S0167404824003237},
author = {Malak Annabi and Abdelhafid Zeroual and Nadhir Messai}
}

@article{HaddadPajouh,
title = {A survey on internet of things security: Requirements, challenges, and solutions},
journal = {Internet of Things},
volume = {14},
pages = {100129},
year = {2021},
issn = {2542-6605},
doi = {https://doi.org/10.1016/j.iot.2019.100129},
url = {https://www.sciencedirect.com/science/article/pii/S2542660519302288},
author = {Hamed HaddadPajouh and Ali Dehghantanha and Reza {M. Parizi} and Mohammed Aledhari and Hadis Karimipour}
}

@ARTICLE{Ratasich,
  author={Ratasich, Denise and Khalid, Faiq and Geissler, Florian and Grosu, Radu and Shafique, Muhammad and Bartocci, Ezio},
  journal={IEEE Access}, 
  title={A Roadmap Toward the Resilient Internet of Things for Cyber-Physical Systems}, 
  year={2019},
  volume={7},
  number={},
  pages={13260-13283},
  doi={10.1109/ACCESS.2019.2891969}
}

@ARTICLE{Tange,
  author={Tange, Koen and De Donno, Michele and Fafoutis, Xenofon and Dragoni, Nicola},
  journal={IEEE Communications Surveys \& Tutorials}, 
  title={A Systematic Survey of Industrial Internet of Things Security: Requirements and Fog Computing Opportunities}, 
  year={2020},
  volume={22},
  number={4},
  pages={2489-2520},
  doi={10.1109/COMST.2020.3011208}
  }

@article{Kriaa,
title = {A survey of approaches combining safety and security for industrial control systems},
journal = {Reliability Engineering \& System Safety},
volume = {139},
pages = {156-178},
year = {2015},
issn = {0951-8320},
doi = {https://doi.org/10.1016/j.ress.2015.02.008},
url = {https://www.sciencedirect.com/science/article/pii/S0951832015000538},
author = {Siwar Kriaa and Ludovic Pietre-Cambacedes and Marc Bouissou and Yoran Halgand}
}

@ARTICLE{Humayed,
  author={Humayed, Abdulmalik and Lin, Jingqiang and Li, Fengjun and Luo, Bo},
  journal={IEEE Internet of Things Journal}, 
  title={Cyber-Physical Systems Security—A Survey}, 
  year={2017},
  volume={4},
  number={6},
  pages={1802-1831},
  doi={10.1109/JIOT.2017.2703172}
  }

@ARTICLE{Giraldo,
  author={Giraldo, Jairo and Sarkar, Esha and Cardenas, Alvaro A. and Maniatakos, Michail and Kantarcioglu, Murat},
  journal={IEEE Design \& Test}, 
  title={Security and Privacy in Cyber-Physical Systems: A Survey of Surveys}, 
  year={2017},
  volume={34},
  number={4},
  pages={7-17},
  doi={10.1109/MDAT.2017.2709310}
  }

@article{Ding,
title = {A survey on security control and attack detection for industrial cyber-physical systems},
journal = {Neurocomputing},
volume = {275},
pages = {1674-1683},
year = {2018},
issn = {0925-2312},
doi = {https://doi.org/10.1016/j.neucom.2017.10.009},
url = {https://www.sciencedirect.com/science/article/pii/S0925231217316351},
author = {Derui Ding and Qing-Long Han and Yang Xiang and Xiaohua Ge and Xian-Ming Zhang}
}

@article{Dibaji,
title = {A systems and control perspective of CPS security},
journal = {Annual Reviews in Control},
volume = {47},
pages = {394-411},
year = {2019},
issn = {1367-5788},
doi = {https://doi.org/10.1016/j.arcontrol.2019.04.011},
url = {https://www.sciencedirect.com/science/article/pii/S1367578819300185},
author = {Seyed Mehran Dibaji and Mohammad Pirani and David Bezalel Flamholz and Anuradha M. Annaswamy and Karl Henrik Johansson and Aranya Chakrabortty}
}

@ARTICLE{Kim,
  author={Kim, Sangjun and Park, Kyung-Joon and Lu, Chenyang},
  journal={IEEE Communications Surveys \& Tutorials}, 
  title={A Survey on Network Security for Cyber–Physical Systems: From Threats to Resilient Design}, 
  year={2022},
  volume={24},
  number={3},
  pages={1534-1573},
  doi={10.1109/COMST.2022.3187531}
  }

@article{Segovia-Ferreira,
author = {Segovia-Ferreira, Mariana and Rubio-Hernan, Jose and Cavalli, Ana and Garcia-Alfaro, Joaquin},
title = {A Survey on Cyber-Resilience Approaches for Cyber-Physical Systems},
year = {2024},
issue_date = {August 2024},
publisher = {Association for Computing Machinery},
address = {New York, NY, USA},
volume = {56},
number = {8},
issn = {0360-0300},
url = {https://doi.org/10.1145/3652953},
doi = {10.1145/3652953},
journal = {ACM Comput. Surv.},
month = apr,
articleno = {202},
numpages = {37}
}

@article{ALABA2017,
  title={Internet of Things security: A survey.},
  author={Alaba, Fadele Ayotunde and Othman, Mazliza and Hashem, Ibrahim Abaker Targio and Alotaibi, Faiz},
  journal={J. Netw. Comput. Appl.},
  volume={88},
  number={March},
  pages={10--28},
  year={2017}
}

@article{FEDERICI2023,
  author  = {Federici, Fabio and Martintoni, Davide and Senni, Valerio},
  title   = {A Zero-Trust Architecture for Remote Access in Industrial
             {IoT} Infrastructures},
  journal = {Electronics},
  volume  = {12},
  number  = {3},
  pages   = {566},
  year    = {2023},
  doi     = {10.3390/electronics12030566}
}

@article{HASSIJA2019,
  author  = {Hassija, Vikas and Chamola, Vinay and Saxena, Vikas
             and Jain, Divyansh and Goyal, Pranav and Sikdar, Biplab},
  title   = {A Survey on {IoT} Security: Application Areas, Security Threats,
             and Solution Architectures},
  journal = {IEEE Access},
  volume  = {7},
  pages   = {82721--82743},
  year    = {2019},
  doi     = {10.1109/ACCESS.2019.2924045}
}

@techreport{KINDERVAG2010,
  author      = {Kindervag, John},
  title       = {No More Chewy Centers: Introducing the Zero Trust Model of
                 Information Security},
  institution = {Forrester Research},
  address     = {Cambridge, MA, USA},
  year        = {2010},
  month       = sep,
  note        = {Published September 14, 2010; updated September 17, 2010}
}

@article{LANGNER2011,
  author  = {Langner, Ralph},
  title   = {Stuxnet: Dissecting a Cyberwarfare Weapon},
  journal = {IEEE Security \& Privacy},
  volume  = {9},
  number  = {3},
  pages   = {49--51},
  year    = {2011},
  doi     = {10.1109/MSP.2011.67}
}

@article{LIU2011,
  author  = {Liu, Yao and Ning, Peng and Reiter, Michael K.},
  title   = {False Data Injection Attacks against State Estimation in
             Electric Power Grids},
  journal = {ACM Transactions on Information and System Security},
  volume  = {14},
  number  = {1},
  pages   = {13:1--13:33},
  articleno = {13},
  year    = {2011},
  doi     = {10.1145/1952982.1952995}
}

@techreport{NIST800207,
  author      = {Rose, Scott and Borchert, Oliver and Mitchell, Stu
                 and Connelly, Sean},
  title       = {Zero Trust Architecture},
  institution = {National Institute of Standards and Technology},
  type        = {NIST Special Publication},
  number      = {800-207},
  address     = {Gaithersburg, MD, USA},
  year        = {2020},
  doi         = {10.6028/NIST.SP.800-207}
}

@article{SYED2022,
  author  = {Syed, Naeem Firdous and Shah, Syed W. and Shaghaghi, Ali
             and Anwar, Adnan and Baig, Zubair and Doss, Robin},
  title   = {Zero Trust Architecture ({ZTA}): A Comprehensive Survey},
  journal = {IEEE Access},
  volume  = {10},
  pages   = {57143--57179},
  year    = {2022},
  doi     = {10.1109/ACCESS.2022.3174679}
}

@article{MATPOWER,
  author  = {Zimmerman, Ray Daniel and Murillo-S{\'a}nchez, Carlos E.
             and Thomas, Robert J.},
  title   = {{MATPOWER}: Steady-State Operations, Planning, and Analysis
             Tools for Power Systems Research and Education},
  journal = {IEEE Transactions on Power Systems},
  volume  = {26},
  number  = {1},
  pages   = {12--19},
  year    = {2011},
  doi     = {10.1109/TPWRS.2010.2051168}
}

@book{ABUR2004,
  author    = {Abur, Ali and Exp{\'o}sito, Antonio G{\'o}mez},
  title     = {Power System State Estimation: Theory and Implementation},
  publisher = {Marcel Dekker},
  address   = {New York, NY, USA},
  year      = {2004},
  isbn      = {9780824755706}
}

@techreport{CISA2023ZTMM,
  author      = {{Cybersecurity and Infrastructure Security Agency}},
  title       = {Zero Trust Maturity Model},
  institution = {Cybersecurity and Infrastructure Security Agency},
  type        = {Version 2.0},
  year        = {2023},
  month       = apr,
  url         = {https://www.cisa.gov/resources-tools/resources/zero-trust-maturity-model}
}

@article{HiGHS2018,
  author  = {Huangfu, Qi and Hall, J. A. J.},
  title   = {Parallelizing the Dual Revised Simplex Method},
  journal = {Mathematical Programming Computation},
  year    = {2018},
  volume  = {10},
  number  = {1},
  pages   = {119--142},
  doi     = {10.1007/s12532-017-0130-5}
}

@inproceedings{MUNIR2024,
  author    = {Munir, Md. Shirajum and Proddatoori, Sravanthi
               and Muralidhara, Manjushree and Saad, Walid
               and Han, Zhu and Shetty, Sachin},
  title     = {A Zero Trust Framework for Realization and Defense
               Against Generative {AI} Attacks in Power Grid},
  booktitle = {2024 IEEE International Conference on Communications
               (ICC)},
  year      = {2024},
  pages     = {2482--2488},
  doi       = {10.1109/ICC51166.2024.10622574}
}

@article{Mosenia2017,
  author  = {Mosenia, Arsalan and Jha, Niraj K.},
  title   = {A Comprehensive Study of Security of Internet-of-Things},
  journal = {IEEE Transactions on Emerging Topics in Computing},
  year    = {2017},
  volume  = {5},
  number  = {4},
  pages   = {586--602},
  doi     = {10.1109/TETC.2016.2606384}
}

@article{Neshenko2019,
  author  = {Neshenko, Nataliia and Bou-Harb, Elias
             and Crichigno, Jorge and Kaddoum, Georges and Ghani, Nasir},
  title   = {Demystifying {IoT} Security: An Exhaustive Survey on
             {IoT} Vulnerabilities and a First Empirical Look on
             Internet-Scale {IoT} Exploitations},
  journal = {IEEE Communications Surveys \& Tutorials},
  year    = {2019},
  volume  = {21},
  number  = {3},
  pages   = {2702--2733},
  doi     = {10.1109/COMST.2019.2910750}
}

@article{Pandapower2018,
  author  = {Thurner, Leon and Scheidler, Alexander
             and Sch{\"a}fer, Florian and Menke, Jan-Hendrik
             and Dollichon, Julian and Meier, Friederike
             and Meinecke, Steffen and Braun, Martin},
  title   = {pandapower---An Open-Source {Python} Tool for Convenient
             Modeling, Analysis, and Optimization of Electric Power Systems},
  journal = {IEEE Transactions on Power Systems},
  year    = {2018},
  volume  = {33},
  number  = {6},
  pages   = {6510--6521},
  doi     = {10.1109/TPWRS.2018.2829021}
}

@misc{RANATHUNGA2026,
  author        = {Ranathunga, Tharindu and Fernando, Kavishka and Rea, Susan},
  title         = {When Agents Control Robots: A Zero Trust Policy Model
                   for Agentic Cyber-Physical Systems},
  year          = {2026},
  eprint        = {2605.25653},
  archivePrefix = {arXiv},
  primaryClass  = {cs.CR},
  url           = {https://arxiv.org/abs/2605.25653}
}

@article{Roman2013,
  author  = {Roman, Rodrigo and Zhou, Jianying and Lopez, Javier},
  title   = {On the Features and Challenges of Security and Privacy
             in Distributed Internet of Things},
  journal = {Computer Networks},
  year    = {2013},
  volume  = {57},
  number  = {10},
  pages   = {2266--2279},
  doi     = {10.1016/j.comnet.2012.12.018}
}

@article{SaltzerSchroeder1975,
  author  = {Saltzer, Jerome H. and Schroeder, Michael D.},
  title   = {The Protection of Information in Computer Systems},
  journal = {Proceedings of the IEEE},
  year    = {1975},
  volume  = {63},
  number  = {9},
  pages   = {1278--1308},
  doi     = {10.1109/PROC.1975.9939}
}

@article{SciPy2020,
  author  = {Virtanen, Pauli and Gommers, Ralf and Oliphant, Travis E.
             and Haberland, Matt and Reddy, Tyler and Cournapeau, David
             and Burovski, Evgeni and Peterson, Pearu and Weckesser, Warren
             and Bright, Jonathan and {van der Walt}, St{\'e}fan J.
             and Brett, Matthew and Wilson, Joshua and Millman, K. Jarrod
             and {van Mulbregt}, Paul and {SciPy 1.0 Contributors}},
  title   = {{{SciPy} 1.0: Fundamental Algorithms for Scientific
             Computing in Python}},
  journal = {Nature Methods},
  year    = {2020},
  volume  = {17},
  number  = {3},
  pages   = {261--272},
  doi     = {10.1038/s41592-019-0686-2}
}

@article{Sicari2015,
  author  = {Sicari, Sabrina and Rizzardi, Alessandra
             and Grieco, Luigi Alfredo and Coen-Porisini, Alberto},
  title   = {Security, Privacy and Trust in Internet of Things:
             The Road Ahead},
  journal = {Computer Networks},
  year    = {2015},
  volume  = {76},
  pages   = {146--164},
  doi     = {10.1016/j.comnet.2014.11.008}
}
